\documentclass[aps,pra,reprint,superscriptaddress,10pt,longbibliography,nofootinbib]{revtex4-1}
\usepackage{color}
\usepackage{verbatim}
\usepackage{amsmath}
\usepackage{amssymb}
\usepackage{wrapfig}
\usepackage{braket}
\usepackage{amsthm}
\usepackage{dsfont}
\usepackage{graphicx}
\usepackage{cancel}

\usepackage{pifont}
\usepackage{lilyglyphs}

\definecolor{blue}{rgb}{0.0, 0.0, 0.5}

\definecolor{Gray}{gray}{0.6}
\definecolor{Red}{RGB}{174,47,42}
\definecolor{Blue}{RGB}{149,178,223}

\usepackage[colorlinks = true, citecolor = blue, linkcolor = blue, urlcolor=blue]{hyperref}
\usepackage[caption=false]{subfig}

\newcommand{\Lcal}{{\mathcal{L}}}
\newcommand{\Scal}{{\mathcal{S}}}
\newcommand{\Ucal}{{\mathcal{U}}}

\newcommand{\Ubf}{\mathbf{U}}
\newcommand{\Sbf}{{\mathbf{S}}}
\newcommand{\Tr}{{\mathrm{Tr}}}

\newcommand{\nmin}{{n_\text{min}}}

\makeatletter
\def\@bibdataout@aps{%
	\immediate\write\@bibdataout{%
		@CONTROL{%
			apsrev41Control%
			\longbibliography@sw{%
				,author="08",editor="1",pages="2",title="0",year="1"%
			}{%
				,author="08",editor="1",pages="2",title="",year="1"%
			}%
		}%
	}%
	\if@filesw \immediate \write \@auxout {\string \citation {apsrev41Control}}\fi 
}
\makeatother

\begin{document}
\title{Quantum jump Monte Carlo simplified: \\Abelian symmetries}

\author{Katarzyna Macieszczak}
\affiliation{TCM Group, Cavendish  Laboratory,  University  of  Cambridge,	J.  J.  Thomson  Ave.,  Cambridge  CB3  0HE,  United Kingdom}
\author{Dominic C. Rose}
\affiliation{School of Physics and Astronomy, University of Nottingham, University Park,  Nottingham NG7 2RD, United Kingdom}
\affiliation{Centre for the Mathematics and Theoretical Physics of Quantum Non-Equilibrium Systems, University of Nottingham, University Park,  Nottingham NG7 2RD, United Kingdom}

\begin{abstract}
	We consider Markovian dynamics of a finitely dimensional open quantum system featuring a weak unitary symmetry, i.e., when the action of a unitary symmetry on the space of density matrices commutes with the master operator governing the dynamics. 
	We show how to encode the weak symmetry in quantum stochastic dynamics of the system by constructing a weakly symmetric representation of the master operator: a symmetric Hamiltonian, and jump operators connecting only the symmetry eigenspaces with a fixed eigenvalue ratio. In turn, this representation simplifies both the construction of the master operator as well as quantum jump Monte Carlo simulations, where,
	for a symmetric initial state, stochastic trajectories of the system state are supported within a single symmetry eigenspace at a time, which is changed only by the action of an asymmetric jump operator. Our results generalize directly to the case of multiple Abelian weak symmetries.
\end{abstract}



\maketitle

\section{Introduction}

Markovian open quantum systems describe a broad class of systems interacting weakly with environments whose dynamics are much faster than those of the system itself, as relevant, e.g., for atomic, molecular and optical physics~\cite{Gordon2006} as well as optomechanics~\cite{Aspelmeyer2014}. 
This leads to system dynamics efficiently described by a local-in-time master equations~\cite{Lindblad1976,Gorini1976}, so that both the dynamics and stationary states can be found by its numerical integration or diagonalization. Since the space on which the corresponding master operator acts scales quadratically with the system dimension, other methods for exact numerical simulations of dynamics have been developed scaling with respect to the system dimension rather than its square,  such as the quantum jump Monte Carlo (QJMC) approach~\cite{Dum1992,Molmer92,Molmer93,Plenio1998,Daley2014}, also known as the quantum trajectory technique or Monte Carlo wave-function method,  which corresponds to the stochastic description of system dynamics in terms of quantum Langevin equations~\cite{Ritsch1988,Gardiner1985,Parkins1988} or continuous measurement theory~\cite{Gardiner2004,Wiseman2010}.

\begin{figure}[t!]
	\begin{center}
		\vspace*{0mm}
		\includegraphics[width=\columnwidth]{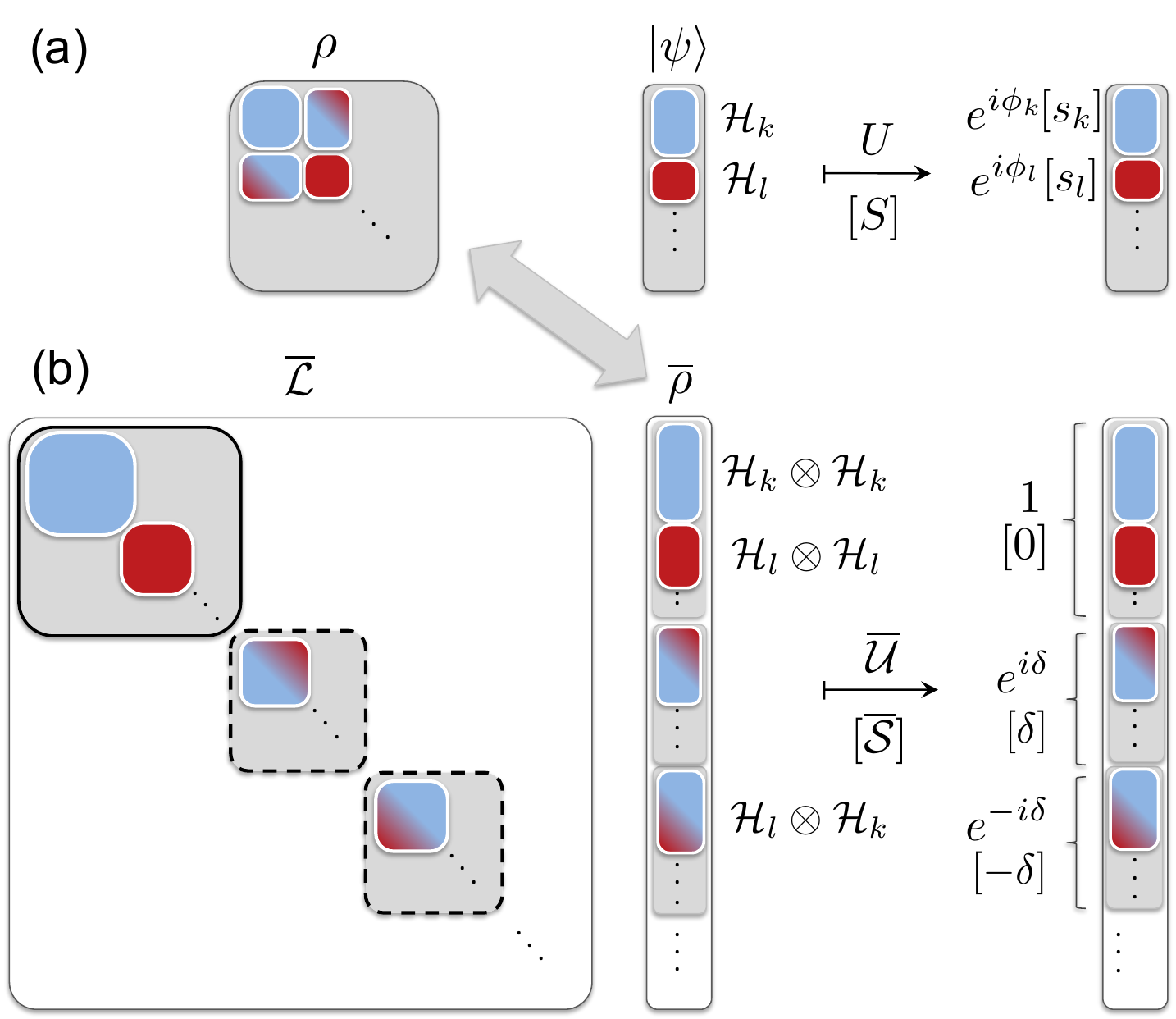} 
		\vspace*{-1mm}
		\caption{
			\textbf{Dynamics with a weak symmetry}. 	\textbf{(a)} Consider a basis of eigenstates of a unitary operator $U$ (or a Hermitian operator $S$) and denote $\mathcal{H}_k$ the eigenspace with $e^{i\phi_k}$ (or $s_k$) eigenvalue. A density matrix $\rho$ features both support inside  $\mathcal{H}_k$ (solid diagonal blocks) and coherences between symmetry eigenspaces (shaded off-diagonal blocks). \textbf{(b)} When the density matrix is	expressed as a vector $\overline{\rho}$, the master operator becomes a matrix $\overline{\Lcal}$ [cf.~Eq.~\eqref{eq:master_lin}]. 
			The weak symmetry implies that the eigenspaces of the symmetry superoperator $\overline{\Ucal}$ (or $\overline{\Scal}$) evolve independently [gray blocks (with black solid or black dashed contour)].
			When ratios of $U$ eigenvalues (or gaps in $S$ eigenvalues) are trivially degenerate [$e^{i(\phi_k-\phi_l)}=e^{i(\phi_{k'}-\phi_{l'})}$ (or $s_k-s_l=s_{k'}-s_{l'}$)  for $k\neq l$ implies $k=k'$ and $l=l'$], individual coherences evolve independently  (gray blocks with black dashed contour vanish while colored blocks remain). 
		}\vspace*{-11mm}
		\label{fig:matrix}
	\end{center}
\end{figure}

Similarly as in closed quantum system dynamics governed by Hamiltonians, the presence of symmetries in master equations is known to simplify the structure of corresponding master operators, although, due to the presence of dissipation, their symmetries are not in general related to conservation laws~\cite{Baumgartner2008a,Buca2012,Albert2014,Gough2015} and their stationary states are unique~\cite{Evans1977,Schirmer2010,Nigro2019,Nigro2020}. In this work, we show how a weak symmetry of the master equation can be encoded in the corresponding stochastic dynamics of an open quantum system: via a symmetric Hamiltonian and jump operators connecting only the symmetry eigenspaces with a fixed eigenvalue ratio, which we refer as a \emph{weakly symmetric representation of a master operator}. This has direct consequences for the numerics:  QJMC simulations are simplified,  particularly for symmetric initial states, which remain symmetric and thus confined to a single symmetry eigenspace at a time. In turn, also the construction of the master operator, which describes change in time of the average system state, is simplified.  Our results carry on directly to the case of multiple weak symmetries, provided their action on density matrices commutes, that is, they correspond to an \emph{Abelian} group. This is  illustrated with a dissipative spin system featuring both a weak translation symmetry and a weak rotation symmetry.

This article is structured as follows. 
In Sec.~\ref{sec:preliminaries} we review Markovian dynamics with weak symmetries.
We define weakly symmetric representations and  show how to construct them in Sec.~\ref{sec:R}. 
We then explain how such representations simplify stochastic dynamics in Sec.~\ref{sec:trajectories}, leading to simplified construction of the master operator and QJMC simulations, as outlined in Sec.~\ref{sec:numerics}. 
Finally, we discuss examples of many-body systems with weak symmetries in Sec.~\ref{sec:examples}.  


\section{Weak unitary symmetries of open quantum system dynamics} \label{sec:preliminaries}

Here, we review Markovian dynamics of open quantum systems featuring weak symmetries.

\subsection{Open quantum system dynamics} \label{sec:Lindblad}

The Markovian dynamics of an open quantum system is governed by a Gorini-Kossakowski-Lindblad-Sudarshan master equation~\cite{Lindblad1976,Gorini1976},
\begin{eqnarray}~\label{eq:master}
\frac{d }{d  t} \rho_t &=& -i[H,\rho_t]+\frac{1}{2}\sum_j\left(2\, J_j \,\rho_t\,J_j^\dagger -J_j^\dagger J_j\rho_t+ \rho_t J_j^\dagger J_j\right)\nonumber\\
&\equiv&\Lcal(\rho_t),
\end{eqnarray}
where $\rho_t$ is a density matrix describing the average  state of the system at time $t$, 
$H$ is a system Hamiltonian (we take $\hbar\equiv1$) and $J_j$ are so-called jump operators describing the interaction with external environments. The dynamics in Eq.~\eqref{eq:master} is completely positive and trace preserving [$\Lcal^\dagger(\mathds{1})=0$], from which it follows that there exists a stationary state of the system [$\Lcal(\rho_\text{ss})=0$]. We  refer to the superoperator $\Lcal$ as the \emph{master operator}. A Hamiltonian and jump operators are not uniquely defined for a given master operator~\cite{Wolf2012}, and we refer to their particular choice  as its \emph{representation}.


\subsection{Weak unitary symmetries} \label{sec:symmetry}

\subsubsection{Definition}

Dynamics of an open quantum system features a dynamical symmetry~\cite{Baumgartner2008a} or a \emph{weak symmetry} of the dynamics~\cite{Buca2012,Albert2014} when it commutes with a unitary transformation of system states. 

For a unitary operator $U$ on the system Hilbert space $\mathcal{H}$, the dynamics features the corresponding weak symmetry when the master operator is symmetric,
 \begin{equation}\label{eq:weakU}
\Ucal\Lcal\,\Ucal^\dagger=\Lcal,
\end{equation}
with respect to the action of the symmetry on density matrices, $\Ucal(\rho)\equiv U \rho U^\dagger$.
Indeed, Eq.~\eqref{eq:weakU}  is equivalent to  $[\Ucal,\Lcal]=0$.
This is often referred to as a \emph{discrete weak symmetry}, as it follows that the dynamics features weak symmetries for unitary operators $U^n$, with $n\in \mathbb{Z}$.

\emph{Abelian weak symmetries} correspond to weak symmetries with commuting symmetry superoperators, $[\Ucal_1,\Ucal_2]=0$, which requires the symmetry operators to commute as well, $[U_1,U_2]=0$. A special case is a \emph{continuous weak symmetry}, that is a weak symmetry for $U_\phi= e^{i\phi S}$, where $S$ is  a Hermitian operator  and $\phi\in\mathbb{R}$, which requires [cf.~Eq.~\eqref{eq:weakU}]
\begin{equation}\label{eq:weakS}
[\Scal,\Lcal]=0,
\end{equation}
with $\Scal(\rho)\equiv[S,\rho]$.

Note that weak symmetries in Eq.~\eqref{eq:weakU} and~\eqref{eq:weakS} are defined as symmetries of the master operator in Eq.~\eqref{eq:master}.
We will discuss the resulting properties of a Hamiltonian and jump operators  in Sec.~\ref{sec:R}, and the structure of the corresponding quantum stochastic dynamics in Sec.~\ref{sec:trajectories}.

\subsubsection{Resulting structure of master operator}\label{sec:L_cons0}

 It is known that weak symmetries limit the structure of the master operator, which can be exploited to simplify its diagonalization or numerical integration required to solve the dynamics of system states~\cite{Albert2014}, as we review below. \\


The structure of the master operator due to a weak symmetry can be conveniently seen  in the Liouville representation (see Fig.~\ref{fig:matrix}).  For
an orthonormal basis $\{|\psi_k\rangle\}_{k=1}^{\dim(\mathcal{H})}$ of the system Hilbert space $\mathcal{H}$, and a density matrix $\rho$ represented as a vector 
$\overline{\rho}\equiv \sum_{k,l=1}^{\dim(\mathcal{H})}\langle \psi_k|\rho |\psi_l\rangle \, |\psi_k\rangle\otimes |\psi_l\rangle\in \mathcal{H}\otimes\mathcal{H}$, its dynamics [cf.~Eq.~\eqref{eq:master}]
\begin{equation}\label{eq:master_lin0}
	\frac{d}{dt}\overline{\rho}_t=\overline{\Lcal}\,\overline{\rho}_t
\end{equation} is governed by the matrix
\begin{eqnarray}\label{eq:master_lin}
	\overline{\Lcal}&\equiv&-i H\otimes \mathds{1}+ i \mathds{1}\otimes H^* \\\nonumber
	&&+ \frac{1}{2}\sum_j\left(2\, J_j \otimes J_j^* -J_j^\dagger J_j\otimes  \mathds{1} -\mathds{1} \otimes J_j^\mathrm{T} J_j^*\right),\\\nonumber
\end{eqnarray}
where $^*$ and $^\mathrm{T}$ denote the complex conjugation and the matrix transposition in the chosen basis, respectively, and we used $H^T=H^*$ since $H=H^\dagger$.

A weak symmetry in Eq.~\eqref{eq:weakU} corresponds to 
\begin{eqnarray}\label{eq:weakU_lin}
[\overline{\Ucal},\overline{\Lcal}]=0,
\end{eqnarray}
where 
$\overline{\Ucal}\equiv U\otimes U^*$.
Therefore the weak symmetry is equivalent to conservation of the eigenspaces of symmetry superoperator $\overline{\Ucal}$,
\begin{equation} \label{eq:weakU_lin2}
	\sum_{\substack{k,l:\\e^{i(\phi_k-\phi_l)}=e^{i\delta}}}  \!\!\!\!\!\!\!\! \overline{\Lcal}     \left(\mathds{1}_{\mathcal{H}_k}  \otimes \mathds{1}_{\mathcal{H}_l}^*\right)= \!\!\!\!\!\!\!\! 	\sum_{\substack{k,l:\\e^{i(\phi_k-\phi_l)}=e^{i\delta}}}    \!\!\!\!\!\!\!\!   \left(\mathds{1}_{\mathcal{H}_k}  \otimes \mathds{1}_{\mathcal{H}_l}^*\right)\overline{\Lcal} ,
\end{equation}
where $\mathcal{H}_k$ is the eigenspaces of $U$ corresponding to an eigenvalue $e^{i\phi_k}$  and  $\mathds{1}_{\mathcal{H}_k}$ is the corresponding orthogonal projection, while $e^{i\delta}$ is an eigenvalue of $\overline{\Ucal}$. In particular, considering $e^{i\delta}=1$  leads to $k=l$ in Eqs.~\eqref{eq:weakU_lin2},  we obtain that the symmetric part of a system state evolves independently from the coherences between symmetry eigenspaces. 
As a result, in a generic case when the stationary state of the dynamics is unique~\cite{Evans1977,Schirmer2010,Nigro2019}, it is symmetric and can be found by solving the dynamics restricted to the symmetric eigenspace~\cite{Hartmann2016,Albert2014,Nigro2020,Cattaneo2020}. Furthermore, averages and higher-order correlations for symmetric system observables can be found, without loss of generality, by solving the dynamics restricted to that eigenspace, that is, by considering symmetric initial states.
Finally, choosing the basis $\{|\psi_k\rangle\}_{k=1}^{\dim(\mathcal{H})}$ as
an eigenbasis of $U$, the matrix $\overline{\Ucal}$ becomes diagonal. Thus, after reordering the basis $\{|\psi_k\rangle\otimes |\psi_l\rangle\}_{k,l=1}^{\dim(\mathcal{H})}$ of $\mathcal{H}\otimes\mathcal{H}$ to group together elements corresponding to the same eigenvalues of $\overline{\Ucal}$, $\overline{\Lcal}$ in Eq.~\eqref{eq:master_lin} becomes \emph{block diagonal},  with the blocks corresponding to the eigenspaces of $\overline{\Ucal}$ (see Fig.~\ref{fig:matrix}). The dynamics of the system states can then be found by diagonalization or numerical integration of individual blocks.\footnote{The preservation of operator hermiticity by the dynamics and symmetry superoperators, $[\Lcal(\rho)]^\dagger=\Lcal(\rho^\dagger)$ and $[\Ucal(\rho)]^\dagger=\Ucal(\rho^\dagger)$, leads to the dynamics of the blocks corresponding to complex-conjugate pairs of symmetry eigenspaces being related by the complex conjugation and swap operation $T$, as $T(\overline{\Lcal}\overline{\rho})^*=\overline{\Lcal}(T\overline{\rho}^*)$, where ${T |\psi_k\rangle\otimes |\psi_l\rangle=|\psi_l\rangle\otimes |\psi_k\rangle}$ for the orthonormal basis $\{|\psi_k\rangle\}_k^{\text{dim}(\mathcal{H})}$.}

Similarly, as superoperators of Abelian symmetries commute in the Liouville representation, $[\overline{\Ucal}_1,\overline{\Ucal}_2]=0$,  the master operator $\overline{\Lcal}$ featuring corresponding weak symmetries conserves all intersections of their eigenspaces. Let orthogonal subspaces $\{\mathcal{H}_k\}_k$ be defined as intersections of eigenspaces of the symmetry operators, so that  each $\mathcal{H}_k$ corresponds to a different set of eigenvalues for the symmetry operators. Then the sums in Eq.~\eqref{eq:weakU_lin2} are replaced by sums over $k,l$ with the ratios of the symmetry operator eigenvalues  corresponding to a given set of eigenvalues for the corresponding symmetry superoperators.
In particular, for a continuous weak symmetry in Eq.~\eqref{eq:weakS} and an eigenspace $\mathcal{H}_k$ of $S$  corresponding to an eigenvalue $s_k$, $e^{i(\phi_k-\phi_l)}=e^{i\delta}$ in Eq.~\eqref{eq:weakU_lin2}  is replaced by $s_k-s_l=\delta$, where $\delta$ is an eigenvalue of $\overline{\Scal}=S\otimes \mathds{1}-\mathds{1}\otimes S^*$, so that $\delta=0$ corresponds to the symmetric part of a system state.

As we show in Sec.~\ref{sec:L_cons}, the Liouville representation of the master operator in the symmetry eigenbasis can be efficiently constructed using a weakly symmetric representation. Even when restricted to symmetric states, however, the master operator acts on the space of dimension $\sum_k \text{dim}(\mathcal{H}_k)^2$, which can inhibit its diagonalization or numerical integration.  Therefore, in Sec.~\ref{sec:MC}, we  instead  focus on   exploiting Abelian weak symmetries to simplify QJMC simulations.

\section{Weakly symmetric representations} \label{sec:R}
Here, we define and construct weakly symmetric representations for any master operator with a weak symmetry by modifying its Hamiltonian and jump operators into a form respecting the symmetry. We also discuss their nonuniqueness.

\subsection{Definition} 
In Sec.~\ref{sec:R_con}, we show that in the presence of a weak symmetry in Eq.~\eqref{eq:weakU}, there exists a \emph{weakly symmetric representation} with a Hamiltonian $\Tilde{H}$ and jump operators $\{\Tilde{J}_j\}_j$ such that
\begin{equation} \label{eq:weakR}
	\Ucal (\Tilde{H}) = \Tilde{H}\quad\text{and}\quad \forall{j} \,\,\,\Ucal(\Tilde{J}_j) = e^{i\delta_j} \Tilde{J}_j.
\end{equation}  
Since the Hamiltonian and the jump operators in Eq.~\eqref{eq:weakR} are eigenmatrices of the symmetry superoperator $\Ucal$, they are supported on the corresponding eigenspaces,
\begin{subequations}
	\label{eq:HJj}
	\begin{align}\label{eq:H}
		\Tilde{H}&=\sum_{k} \, \mathds{1}_{\mathcal{H}_k} \Tilde{H} \,\mathds{1}_{\mathcal{H}_k},\\
		\label{eq:Jj}
		\Tilde{J}_j&=\!\!\!\!\!\!\!\! \sum_{\substack{k,l:\\e^{i(\phi_k-\phi_l)}=e^{i\delta_j}}} \!\!\!\!\!\!\!\!  \mathds{1}_{\mathcal{H}_k} \Tilde{J}_j \mathds{1}_{\mathcal{H}_l}.
	\end{align}
\end{subequations}
for all $j$. This property plays a crucial role in simplifying numerical simulations of the system dynamics (see Sec.~\ref{sec:numerics}).  Note that when $e^{i\delta_j}=1$ for all $j$, a Hamiltonian and all jump operators themselves are symmetric. This then holds for any representation of the master operator~\cite{Wolf2012} and is known as
a \emph{strong symmetry}~\cite{Buca2012,Albert2014}. Although a strong symmetry implies the corresponding weak symmetry, the converse is not true~\cite{Baumgartner2008a,Buca2012,Albert2014}, as evident by considering weakly symmetric representations in Eqs.~\eqref{eq:HJj}.

Similarly, for Abelian weak symmetries, a Hamiltonian can be chosen symmetric with respect to all symmetry superoperators and jump operators can be chosen as their simultaneous eigenmatrices.
In particular, in the presence of a continuous weak symmetry in Eq.~\eqref{eq:weakS}, there exists a weakly symmetric representation such that
\begin{equation}\label{eq:weakRC}
	\Scal(\Tilde{H})= 0, \quad\Scal(\Tilde{J}_j) = \delta_j \Tilde{J}_j,
\end{equation} 
with $\Tilde{H}$ and $\Tilde{J}_j$ supported as in Eq.~\eqref{eq:HJj}, but with $e^{i(\phi_k-\phi_l)}=e^{i\delta_j}$  replaced by $s_k-s_l=\delta_j$, where $s_k$ is an eigenvalue of $S$ corresponding to an eigenspace $\mathcal{H}_k$.

\subsection{Construction} \label{sec:R_con}

We now give two constructions of weakly symmetric representations from a given representation of the master operator  in Eq.~\eqref{eq:master}, that is, a Hamiltonian $H$ and a set of jump operators $\{J_j\}_j$.
In the first construction, the Hamiltonian is projected on the symmetric eigenspace of the symmetry superoperator, while the jump operators are projected on all its eigenspaces, so that their number $n$  in general increases  to $m$-fold, where $m$ is the number of distinct eigenvalues of the symmetry superoperator. Here the knowledge of symmetry eigenspaces is assumed, which generally requires diagonalizing  the $\text{dim}(\mathcal{H})\times \text{dim}(\mathcal{H})$ matrix of the symmetry operator $U$ (see Fig.~\ref{fig:construction_DD}).
In the second construction, the number of jump operators does not increase as a weakly symmetric representation with the minimal number of jump operators is constructed, but at the cost of diagonalizing two matrices of size $n\times n$ (see Fig.~\ref{fig:construction_min}).

\subsubsection{Weakly symmetric representation by dynamical decoupling} \label{sec:R_con_DD}

\begin{figure}[t!]
 	\begin{center}
 		\centering
 		\includegraphics[width=0.8375\columnwidth]{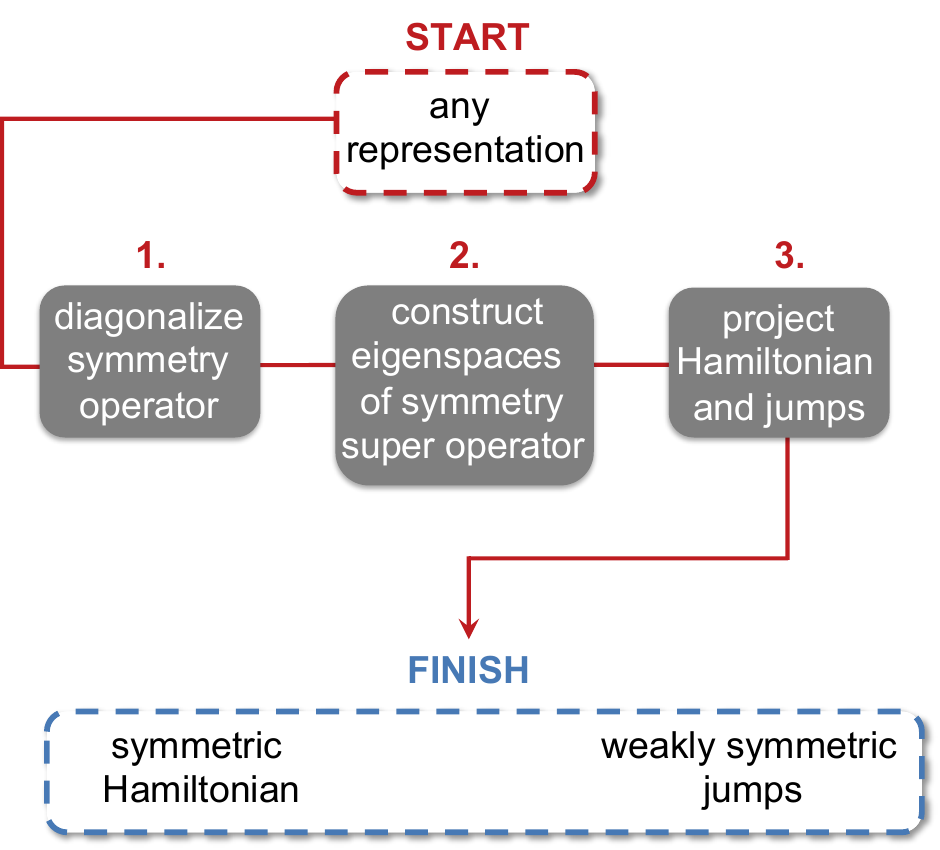}
 		\caption{\textbf{Constructing a weakly symmetric representation by dynamical decoupling}. From any representation a weakly symmetric representation can be obtained by projecting the Hamiltonian on the symmetric eigenspace of the symmetry superoperator and projecting the jump operators on all its eigenspaces; see Eqs.~\eqref{eq:HJjDD}.
 		}\vspace*{-5mm}
 		\label{fig:construction_DD}
 	\end{center}
 \end{figure}

In this construction we use the fact that for dynamics with a weak symmetry in Eq.~\eqref{eq:weakU}
\begin{equation}\label{eq:DD_U}
\Lcal=\lim_{N\rightarrow\infty}\frac{1}{N}\sum_{n=0}^N \Ucal^{n}\Lcal\,\Ucal^{\dagger n}.
\end{equation}
 The right-hand-side limits in Eqs.~\eqref{eq:DD_U}  corresponds to the projection of the master operator on the symmetric part under its transformations $\mathcal{U}(\cdot)\mathcal{U}^\dagger$, and in the construction we consider this projection applied to individual terms appearing in the master equation in Eq.~\eqref{eq:master} (see Fig.~\ref{fig:construction_DD}). Note that such a projection occurs when the dynamics is in fact composed of the system master dynamics governed by $\mathcal{L}$ and much faster unitary dynamics corresponding to $\Ucal$, in which case the former acts as a perturbation of the latter (see Supplemental Material of Refs.~\cite{Macieszczak2016} and~\cite{Burgarth2019}). Therefore, a weak symmetry can be facilitated by \emph{dynamical decoupling}~\cite{Zanardi1999,Viola1999} at a rate much faster than system dynamics but much slower than relaxation of the environment (see Refs.~\cite{Arenz2015,Gough2017}). \\

 \emph{Step 1}. A symmetry operator $U$ is diagonalized  to find its eigenspaces.
 
 \emph{Step 2.} Eigenspaces of the symmetry superoperator $\Ucal$ are constructed.
 
 \emph{Step 3}. The Hamiltonian and jump operators are replaced by  their projections on $\Ucal$ eigenspaces  [cf.~Eq.~\eqref{eq:HJj}]
\begin{subequations}
\label{eq:HJjDD}
\begin{align}\label{eq:HDD}
\Tilde{H}&\equiv\sum_{k}  \mathds{1}_{\mathcal{H}_k} H \,\mathds{1}_{\mathcal{H}_k},\\
\label{eq:JjDD}
\forall_j \forall_{e^{i\delta}}\,\,\,\Tilde{J}_{j,e^{i\delta}}&\equiv\!\!\!\!\!\!\!\! \sum_{\substack{k,l:\\e^{i(\phi_k-\phi_l)}=e^{i\delta}}} \!\!\!\!\!\!\!\!  \mathds{1}_{\mathcal{H}_k} J_j \mathds{1}_{\mathcal{H}_l}.
\end{align}
\end{subequations}
 with  $e^{i\phi_k}$ denoting an eigenvalue of $U$ corresponding to an eigenspace $\mathcal{H}_k$  and  $e^{i\delta}$ denoting an eigenvalue of $\Ucal$.\\

\emph{Construction for Abelian weak symmetries}. Considering Eq.~\eqref{eq:DD_U} for all Abelian weak symmetries present, Eq.~\eqref{eq:HDD} holds with subspaces $\mathcal{H}_k$  defined as intersections of eigenspaces of the symmetry operators, while the sum in Eq.~\eqref{eq:JjDD} is replaced $k,l$ with the ratios of the symmetry operators eigenvalues  corresponding to a given set of eigenvalues for the corresponding symmetry superoperators.
In particular, for a continuous symmetry in Eq.~\eqref{eq:weakS},  
and an eigenspace $\mathcal{H}_k$ of $S$ corresponding to an eigenvalue $s_k$, the Hamiltonian is constructed as in Eq.~\eqref{eq:HDD}, while a jump operator $J_j$ is replaced by the set $\{\Tilde{J}_{j,\delta}\equiv\sum_{k,l:\,s_k-s_l=\delta}  \mathds{1}_{\mathcal{H}_k} J_j \mathds{1}_{\mathcal{H}_l}\}_\delta$ defined for all eigenvalues $\delta$ of $\Scal$ [cf.~Eq.~\eqref{eq:JjDD}].
\\

 \emph{Proof of Eq.~\eqref{eq:HJjDD}}. First, note that $\Ucal\,\Lcal\,\Ucal^\dagger$ corresponds to the master operator with the Hamiltonian $\Ucal(H)$ and the  jump operators  $\Ucal(J_j)$, as seen, for example, in the Liouville representation. The master operator $\overline{\Lcal}$ in Eq.~\eqref{eq:master_lin} is linear in $H\otimes \mathds{1}$ and $ J_j^\dagger J_j\otimes \mathds{1}$, so that in the limit of  the right-hand side in Eq.~\eqref{eq:DD_U} they are replaced by their projection on the symmetric subspace of $\Ucal\otimes \mathcal{I}$, where $\mathcal{I}$ is an identity superoperator, that is, $\Tilde{H}\otimes \mathds{1}$ of Eq.~\eqref{eq:HDD} and $\sum_{k}  \mathds{1}_{\mathcal{H}_k}J_j^\dagger J_j^\dagger  \,\mathds{1}_{\mathcal{H}_k}\otimes \mathds{1}=\sum_{e^{i\delta}} \Tilde{J}_{j,e^{i\delta}}^\dagger \Tilde{J}_{j,e^{i\delta}}^\dagger\otimes \mathds{1}$. Similarly,  the term $J_j\otimes J_j^*$ is projected on the symmetric subspace of $\Ucal\otimes \mathcal{U}^*$ and is thus replaced by $\sum_{e^{i\delta}} \Tilde{J}_{j,e^{i\delta}}\otimes \Tilde{J}_{j,e^{i\delta}}^*$.

\subsubsection{Minimal weakly symmetric representation}\label{sec:R_con_min}

\begin{figure}[t!]
	\begin{center}
		\centering
		\includegraphics[width=0.95\columnwidth]{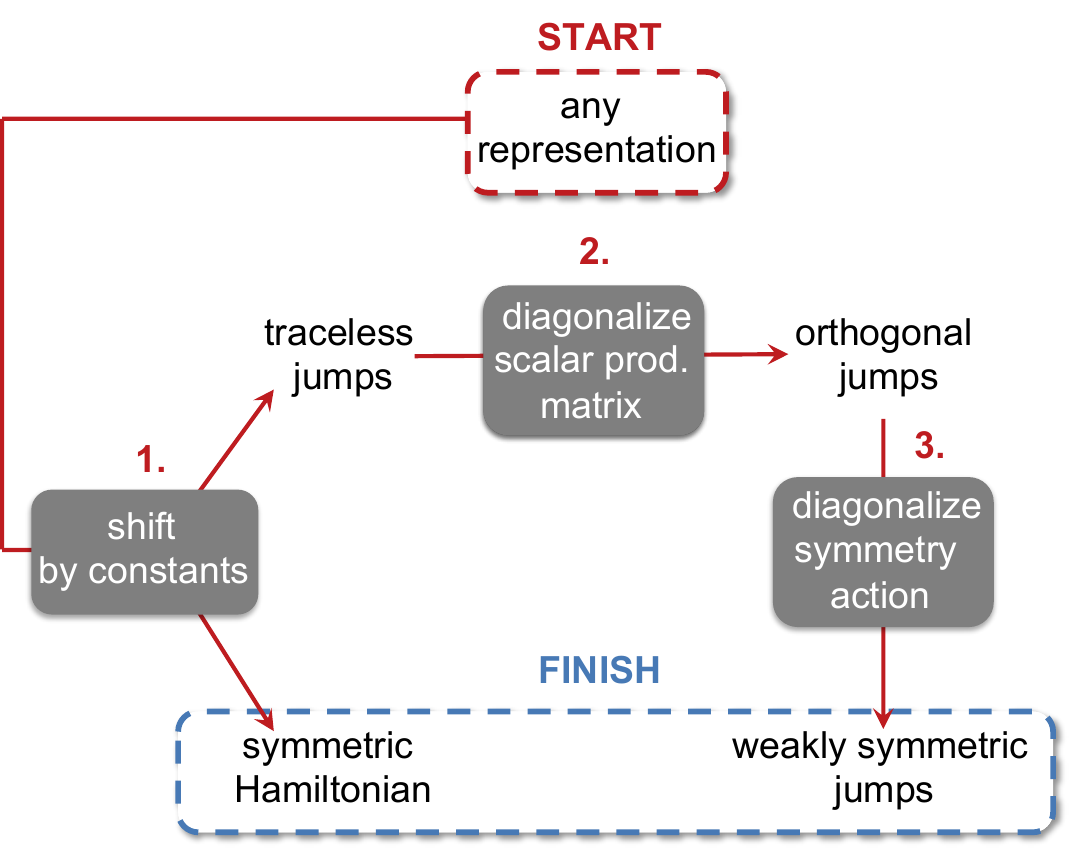}
		\caption{\textbf{Constructing a minimal weakly symmetric representation}. From any representation a minimal weakly symmetric representation can be obtained in three steps: 
			\textbf{1.}~Shifting jump operator to traceless operators [Eq.~\eqref{eq:J'}]  defines a symmetric Hamiltonian [Eq.~\eqref{eq:Htilde}]. 
			\textbf{2.}~Diagonalizing the matrix of scalar products between the traceless jump operators [Eq.~\eqref{eq:C}] yields orthogonal jump operators [Eq.~\eqref{eq:JC}].  
			\textbf{3.}~Diagonalizing the action of the symmetry on the orthogonal jump operators [Eqs.~\eqref{eq:Ubf}] gives weakly symmetric jump operators [Eqs.~\eqref{eq:JtildeU}]. 
		}\vspace*{-5mm}
		\label{fig:construction_min}
	\end{center}
\end{figure}

The steps of this construction are motivated by following two facts (see Fig.~\ref{fig:construction_min}). First, a representation of the master operator with the traceless Hamiltonian and orthogonal traceless jump operators is uniquely defined (up to degeneracy in jump rates) and corresponds to the minimal number, $\nmin\leq n$, of jump operators (see Ref.~\cite{Wolf2012}). Second, the set $\Ucal(H)$, $\{\Ucal(J_{j})\}_{j}$ is a representation of $\Ucal \Lcal\, \Ucal^\dagger$~\cite{Avron2012}, and thus, in the presence of the weak symmetry, it is also a representation of $\Lcal$~\cite{Molmer93}. Since $\Ucal$  does not change the orthogonality and trace of the operator being transformed, we have that in the presence of weak symmetry, the traceless Hamiltonian is necessarily symmetric, while orthogonal traceless jump operators with the same rate are transformed unitarily by $\Ucal$, and thus they can be chosen as its eigenmatrices.   \\

\emph{Step 1}.  Traceless jump operators are constructed by introducing
\begin{equation} \label{eq:J'}
J_j'\equiv J_j-\Tr (J_j)\mathds{1}, \quad j=1,...,n,
\end{equation}
while the Hamiltonian is replaced by  
\begin{equation} \label{eq:Htilde}
\Tilde{H}\equiv H + \frac{i}{2} \sum_j\left[\Tr (J_j^\dagger)\, J_j -\Tr (J_j) J_j^\dagger\right]\!,
\end{equation}
in order to leave the master operator in Eq.~\eqref{eq:master} unchanged [a further shift of the Hamiltonian by $-\Tr(H)\mathds{1}$ only introduces a global phase, and is also symmetric, and thus will be omitted]. 

\emph{Step 2}. A Hermitian matrix of the scalar products,\footnote{The matrix $\mathbf{C}$ is Hermitian and positive semi-definite, as $\mathbf{c}^\dagger \mathbf{C} \mathbf{c} = \Tr (J_c'^\dagger J'_c) \geq 0$, where $J'_c=\sum_{j=1}^n (\mathbf{c})_j J'_j$ is a linear combination of jump operators.}
\begin{equation}\label{eq:C}
(\mathbf{C})_{jk} \equiv\Tr [(J'_j) ^\dagger J'_k], \quad j,k=1,...,n,
\end{equation}
is diagonalized in order to define via its orthonormal eigenvectors, $\mathbf{C} \mathbf{c}_j= \lambda_j \mathbf{c}_j$, $\mathbf{c}_j^\dagger \mathbf{c}_k =\delta_{jk}$ and $\mathbf{c}_j^\dagger \mathbf{c}_k =\delta_{jk}$,
orthogonal jump operators 
\begin{equation} \label{eq:JC}
J_{j}''\equiv\sum_{k=1}^n (\mathbf{c}_j)_k \, J'_k, 
\end{equation}
with the rate determined by the corresponding eigenvalue 
\begin{equation}\label{eq:Lambda}
\Tr [(J_{j}'')^\dagger J_{k}'']= \mathbf{c}_j^\dagger\mathbf{C} \mathbf{c}_k = \lambda_j\, \delta_{jk}\equiv\mathbf{\Lambda}_{jk}.
\end{equation}
We reorder jump operators with decreasing $\lambda_j$ and neglect $j$ with $\lambda_j=0$, as $J_j''=0$ ($j\geq \nmin$).

\emph{Step 3}.  For a weak symmetry in Eq.~\eqref{eq:weakU}, the Hamiltonian in Eq.~\eqref{eq:Htilde} is symmetric,
\begin{equation}\label{eq:HtildeU}
\Ucal( \Tilde{H})=\Tilde{H},
\end{equation} 
while the set of orthogonal jump operators in Eq.~\eqref{eq:JC} is transformed as $\Ucal (J''_{k})= \sum_{j=1}^{\nmin} (\Ubf)_{jk} J_{j}''$, where
\begin{equation}\label{eq:Ubf}
( \Ubf)_{jk} \equiv {\lambda_j}^{-1}\, \Tr    [J_{j}''^\dagger \,\Ucal (J''_{k}) ], \quad j,k=1,...,\nmin
\end{equation}
is a unitary matrix\footnote{$\Ubf$ is unitary as $\Ubf^\dagger$ corresponds to the superoperator $\Ucal^\dagger$,  $(\Ubf^\dagger)_{jk}=(\Ubf)^*_{kj}=\lambda_k^{-1} \Tr    [{J_{k}''}^\dagger\Ucal({J_{j}''}) ]^*=\lambda_k^{-1} \Tr    [\Ucal^\dagger ({J_{k}''}^\dagger)\,{J_{j}''} ]^*=\lambda_k^{-1} \Tr    [{J_{j}''}^\dagger\Ucal^\dagger (J_{k}'') ]=\lambda_j^{-1} \Tr    [{J_{j}''}^\dagger\Ucal^\dagger (J_{k}'') ]$, and thus, from  ${\Ucal\Ucal^\dagger=\Ucal^\dagger \Ucal=\mathcal{I}}$ and Eq.~\eqref{eq:Lambda}, it follows that ${\Ubf\Ubf^\dagger=\Ubf^\dagger \Ubf=\mathbf{I}}$.} which is block diagonal in the eigenspaces of $\mathbf{\Lambda}$. 
The jump operators determined by the  orthonormal eigenvectors of $\Ubf$,   $\Ubf \,\mathbf{u}_j = e^{i\delta_{j}} \mathbf{u}_j$, 
\begin{eqnarray} \label{eq:JtildeU}
\Tilde{J}_{j}\equiv\sum_{k=1}^{\nmin}(\mathbf{u}_j)_k  J''_{k}= \sum_{l=1}^n \sum_{k=1}^{\nmin} (\mathbf{u}_j)_k   (\mathbf{c}_k)_l [J_l-\Tr (J_l)\mathds{1}], \qquad
\end{eqnarray}
 are eigenmatrices of the symmetry superoperator,
\begin{equation}\label{eq:JtildeU2}
\Ucal(\Tilde{J}_{j}) = e^{i\delta_{j}} \Tilde{J}_{j},
\end{equation}
$j=1,...,n_{\min}$. In particular, when diagonalizing blocks in $\Ubf$, the jump operators in Eq.~\eqref{eq:JtildeU} are chosen  orthogonal, as $\Tr (\Tilde{J}_{j}^\dagger \Tilde{J}_{k} )=\mathbf{u}_j^\dagger \mathbf{\Lambda}\mathbf{u}_k$, $j,k=1,...,{\nmin}$.\\

%
%

\emph{Construction for Abelian weak symmetries}. The choice of the Hamiltonian in Eq.~\eqref{eq:Htilde} is independent from the presence of weak symmetries and thus is always symmetric. When weak unitary symmetries commute, $[\Ucal_1,\Ucal_2]=0$, so do the corresponding unitary transformations on the set of orthogonal traceless jump operators [Eq.~\eqref{eq:Ubf}], $[\Ubf_1,\Ubf_2]=0$. Therefore, the jump operators in Eq.~\eqref{eq:JtildeU} can be chosen as eigenmatrices of all symmetry superoperators. 
In particular, for a continuous weak symmetry in Eq.~\eqref{eq:weakS}, the jump operators in Eq.~\eqref{eq:JtildeU} can be defined with orthonormal eigenvectors of a Hermitian matrix\footnote{$\Sbf$ is Hermitian as $(\Sbf)^*_{kj}=\lambda_k^{-1}\Tr    ({J_{k}''}^\dagger [S,J_{j}''] )^*=\lambda_k^{-1}\Tr    ([{J_{j}''}^\dagger,S] J_{k}'')=\lambda_k^{-1}  \Tr    ({J_{j}''}^\dagger\,[S, J_{k}''] )=\lambda_j^{-1}  \Tr    ({J_{j}''}^\dagger\,[S, J_{k}''] )=(\Sbf)_{jk}$.}
\begin{equation}\label{eq:Sbf}
	( \Sbf)_{jk} \equiv  {\lambda_j}^{-1}\, \Tr    [J_{j}''^\dagger \,\Scal (J''_{k}) ], \quad j,k=1,...,\nmin,
\end{equation}
which generates the unitary transformation $\Ubf_\phi=e^{i\phi \Sbf}$ of orthogonal jump operators under $\Ucal_\phi=e^{i\phi \Scal}$ [cf.~Eq.~\eqref{eq:Ubf}] and is block diagonal in the eigenspaces of $\mathbf{\Lambda}$ [Eq.~\eqref{eq:Lambda}].

\subsection{Non-uniqueness} \label{sec:R_non}
In Sec.~\ref{sec:R_con} we constructed two generally different representations of the master equation, which proves that a weakly symmetric representation [Eq.~\eqref{eq:weakR}] is \emph{nonunique}. Here, we characterize the freedom in the choice of weakly symmetric representations.

A general weakly symmetric representation with traceless jump operators is described by an $n\times \nmin$  isometry $\mathbf{V}$ [$(\mathbf{V}^\dagger\mathbf{V})_{jk}=\delta_{jk}$] that does not mix $\Ubf$ eigenspaces [$(\mathbf{V})_{jk}\neq 0$ and $(\mathbf{V})_{jl}\neq 0$ take place only for $e^{i\delta_k}=e^{i\delta_l}$], that is, the set of jump operators $\{\tilde{J}_j\}_j$ defined in Eq.~\eqref{eq:JtildeU} is replaced by $\{\sum_{k=1}^{\nmin}(\mathbf{V})_{jk} \tilde{J}_k\}_{j=1}^n$ (cf.~Ref.~\cite{Wolf2012}). In that case, jump operators are generally not orthogonal, with the scalar product between the $j$th and $k$th jump given by $ (\mathbf{V}^*\mathbf{\Lambda}_\Ubf\mathbf{ V^\text{T}})_{jk}$, with $(\mathbf{\Lambda}_\Ubf)_{jk}=\mathbf{u}_j^\dagger\mathbf{\Lambda}\mathbf{u}_k$ [cf.~Eqs.~\eqref{eq:C} and~\eqref{eq:Lambda}].

A general weakly symmetric representation features to shifted symmetric jump operators (cf.~Ref.~\cite{Wolf2012}). That is, 
symmetric jump operators in a weakly symmetric representation need not to be traceless, as can be shifted by a constant since $[U,\mathds{1}]=0$. Indeed, any jump operator $\Tilde{J}_j$ with $e^{i\delta_j}=1$ in a weakly symmetric representation can be replaced by $\Tilde{J}_j+a_j\mathds{1}$ with  $a_j\in \mathbb{C}$, while the symmetric Hamiltonian $\Tilde{H}$ is transformed to  $\Tilde{H}+b\mathds{1}- i\sum_{j:\,e^{i\delta_j}=1}(a_j^* \Tilde{J}_j-a_j \Tilde{J}_j^\dagger)/2$ with $b\in\mathbb{R}$ [cf.~Eqs.~\eqref{eq:J'} and~\eqref{eq:Htilde}].

\section{Quantum trajectories with weakly symmetric representations} \label{sec:trajectories}
We now briefly discuss implications of the presence of a weak symmetry for the structure of quantum trajectories and the survival of coherences between symmetry eigenspaces.  This structure simplifies the construction of the master operator and reduces both the memory and processing required for QJMC simulations, as we explain in Sec.~\ref{sec:numerics}.

\subsection{Quantum trajectories}

The dynamics in Eq.~\eqref{eq:master} for the system initially in a pure state, $\rho_0=|\psi_0\rangle\!\langle\psi_0|$, can be unraveled as~\cite{Wiseman2010,Gardiner2004}
\begin{eqnarray}\label{eq:rho_series}
	\rho_t
	&=&\sum_{n=0}^\infty \sum_{j_1,...,j_n} \int_{0\leq t_1\leq ...\leq t_n \leq t} d  t_1  \cdots  dt_n\nonumber\\
	&& \qquad|\psi_t (t_1, j_1;...;t_n,j_n) \rangle\!\langle \psi_t (t_1, j_1;...;t_n,j_n) |,\qquad
\end{eqnarray}
with
\begin{subequations}
	\label{eq:psi_series}
	\begin{align}
		\label{eq:psi_t}
		&|\psi_t (t_1, j_1;...;t_n,j_n) \rangle\equiv  \\\nonumber
		&\qquad e^{-i (t-t_n) H_\text{eff}} J_{j_n} \cdots e^{-i (t_2-t_{1}) H_\text{eff}} J_{j_1} |\psi_{t_1}\rangle,\qquad\\
		\label{eq:psi_t0}
		&|\psi_t \rangle\equiv e^{-i t H_\text{eff}}|\psi_0\rangle,
	\end{align}
\end{subequations}
and the effective Hamiltonian 
\begin{equation}\label{eq:Heff}
	H_\text{eff}\equiv H -\frac{i}{2}\sum_j J_j^\dagger J_j.
\end{equation}
%
Equation~\eqref{eq:psi_series} as a function of time is referred to as a (unnormalized) \emph{quantum trajectory}, which at time $t$ describes the (unnormalized) state of the system conditioned on the occurrence of jumps ${j_1}$, ..., ${j_n}$ at respective times $t_1$, ..., $t_n$ [Eq.~\eqref{eq:psi_t}] or their absence [Eq.~\eqref{eq:psi_t0}], which takes place with the probability density and the probability
	\begin{subequations}
\label{eq:p_t}
		\begin{align}
	&p_t(t_1, j_1;...;t_n,j_n)\\\nonumber
	&\qquad\equiv\langle \psi_t (t_1, j_1;...;t_n,j_n)|\psi_t (t_1, j_1;...;t_n,j_n) \rangle,\\
	&p_t\equiv\langle \psi_t |\psi_t  \rangle, \label{eq:p_t0}
	\end{align}
\end{subequations}
respectively.

For dynamics with a unique stationary state, normalized quantum trajectories are  \emph{ergodic}~\cite{Kummerer2004}, 
\begin{equation} \label{eq:ergodic}
	\lim_{T\rightarrow\infty}\frac{1}{T} \int_{0}^{T}d  t\, \frac{| \psi_t (t_1, j_1;...) \rangle\!\langle \psi_t (t_1, j_1;...)|}{p_t(t_1, j_1;...)}=\rho_\text{ss},
\end{equation}
with probability $1$.

\subsection{Simplified quantum trajectories}
We now discuss how quantum trajectories simplify for a weakly symmetric representation (see Fig.~\ref{fig:MC}).

\subsubsection{Symmetric initial states} \label{sec:trajectories_sym}
\begin{figure}[t]
	\begin{center}
		\includegraphics[width=\columnwidth]{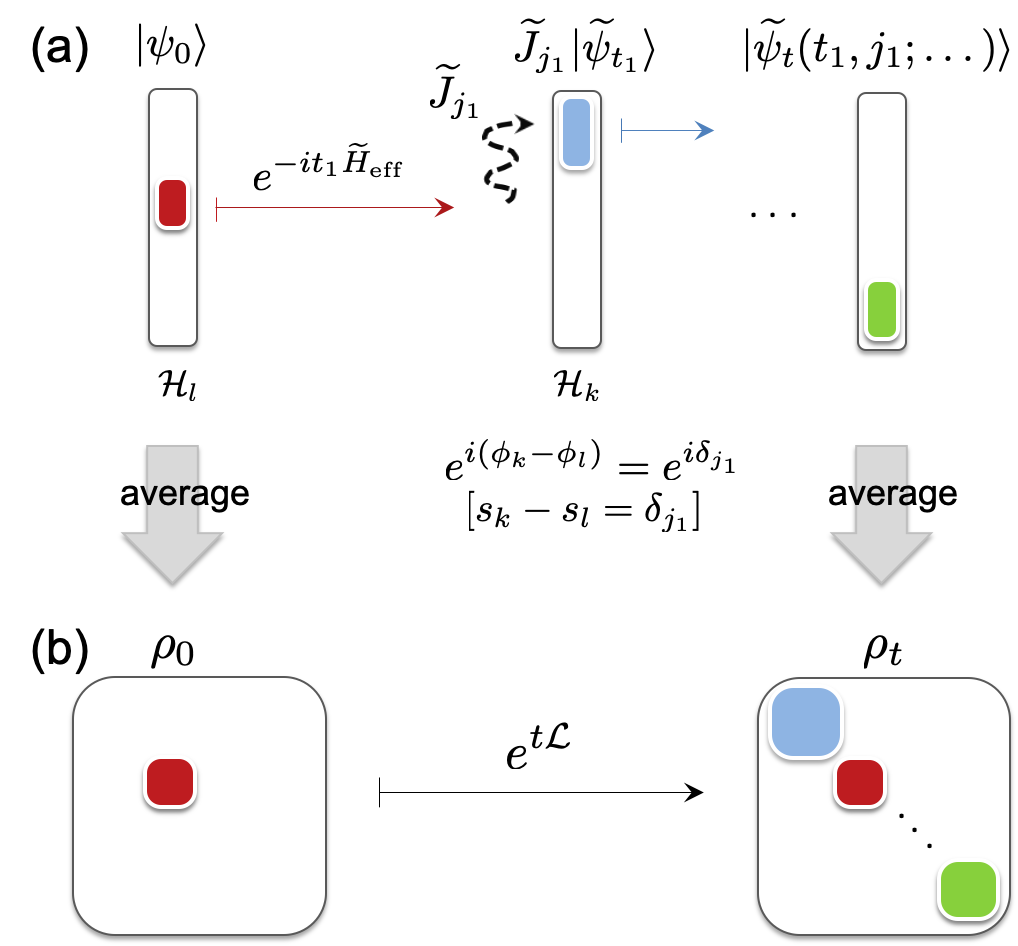} 
		\caption{
			\textbf{Quantum trajectories in a weakly symmetric representation}. \textbf{(a)} An example of a quantum trajectory for a symmetric initial state (red). The system remains in  the original symmetry eigenspace $\mathcal{H}_l$  at least until the first jump $\Tilde{J}_{j_1}$, as the effective Hamiltonian governing no-jump evolution is symmetric  (red horizontal arrow); cf.~Eq~\eqref{eq:Upsi_t}. When the first jump $\Tilde{J}_{j_1}$ is asymmetric, the system is transformed (dashed black arrow)  to another eigenspace 	$\mathcal{H}_k$ (blue) determined by the corresponding eigenvalue $e^{i\delta_{j_1}}$ of $\Ucal$ [or $\delta_{j_1}$ of $\Scal$]; cf.~Eq.~\eqref{eq:UJpsi_t}. The system state remains in $\mathcal{H}_k$ until the next asymmetric jump [see also Figs.~\ref{fig:example_trajectories}(a) and~\ref{fig:example_trajectories}(b)). 
			\textbf{(b)} The average system state [Eq.~\eqref{eq:rho_series}]  is symmetric, $\Ucal(\rho_t)=\rho_t$, since no coherences between symmetry eigenspaces are present in individual trajectories [cf.~Eq.~\eqref{eq:Upsi_t2}]. Its dynamics is governed by the master operator $\Lcal$ (black horizontal arrow) restricted to the symmetric eigenspace of $\Ucal$ (or $\Scal$); cf.~Fig.~\ref{fig:matrix}(b). 
		}\vspace*{-5mm}
		\label{fig:MC}
	\end{center}
\end{figure}

Consider the system initially supported in a symmetry eigenspace $\mathcal{H}_l$, $U|\psi_{0}\rangle=e^{i\phi_l}|\psi_{0}\rangle$, which we refer to as a symmetric initial state since $\Ucal(|\psi_{0}\rangle\!\langle \psi_0|)=0$. For a weakly symmetric representation, the system state remains supported in the same eigenspace when no jumps take place [cf.~Eq.~\eqref{eq:psi_t0} and see Fig.~\ref{fig:MC}(a)], 
\begin{equation}\label{eq:Upsi_t}
	U |\Tilde\psi_t\rangle  =  U e^{-i t \Tilde{H}_\text{eff}} |\psi_{0}\rangle=e^{-i t \Tilde{H}_\text{eff}} U|\psi_{0}\rangle = e^{i\phi_l}|\Tilde\psi_t\rangle,
\end{equation}
because of the symmetry of the effective Hamiltonian,
\begin{equation}\label{eq:HeffU}
	\Ucal(\Tilde{H}_\text{eff})=\Ucal(\Tilde{H})-\frac{i}{2} \sum_j\Ucal(\Tilde{J}_j^\dagger)\,\Ucal(\Tilde{J}_j)=\Tilde{H}_\text{eff}
\end{equation}
from $\Ucal(\Tilde{J}^\dagger_{j})=[\Ucal(\Tilde{J}_{j})]^\dagger= e^{-i\delta_j}\Tilde{J}_{j}^\dagger$.
Occurrence of the first jump $\Tilde{J}_{j_1}$ at time $t_1$ transforms the state $|\Tilde\psi_{t_1}\rangle$ from  $\mathcal{H}_l$ into the symmetry eigenspace $\mathcal{H}_k$ with the eigenvalue $e^{i\phi_k}= e^{i(\delta_{j_1}+\phi_l)}$ [cf.~Eqs.~\eqref{eq:Jj} and~\eqref{eq:psi_t}], 
\begin{equation}\label{eq:UJpsi_t}
	U \Tilde{J}_{j_1} |\Tilde\psi_{t_1}\rangle  =U \Tilde{J}_{j_1} U^\dagger U|\Tilde\psi_{t_1}\rangle =e^{i\delta_{j_1}} \Tilde{J}_{j_1}e^{i\phi_l}|\Tilde\psi_{t_1}\rangle.
\end{equation}
Only for a symmetric jump, $e^{i\delta_{j_1}}=1$, the symmetry eigenspace remains unchanged, $\mathcal{H}_k=\mathcal{H}_l$. The system state remains in $\mathcal{H}_k$ until the next asymmetric jump [see Fig.~\ref{fig:MC}(a)].

Therefore, for an initially symmetric system state, quantum trajectories with a weakly symmetric representation are symmetric at all times,
\begin{equation}\label{eq:Upsi_t2}
	\Ucal\big[|\Tilde\psi_t(...)\rangle\!\langle\Tilde\psi_t(...)|\big]=|\Tilde\psi_t(...)\rangle\!\langle\Tilde\psi_t(...)|.
\end{equation}
In contrast, in a general representation, coherences between symmetry eigenspaces are present in individual quantum trajectories but interfere destructively in the average of Eq.~\eqref{eq:rho_series}. This illustrates the fact that \emph{quantum stochastic dynamics with a weakly symmetric representation features the weak symmetry}, as do the dynamics of a density matrix  under the effective Hamiltonian or the action of any of jump operators [cf.~Eqs.~\eqref{eq:Upsi_t} and~\eqref{eq:UJpsi_t}]. The weak symmetry is then inherited by the average dynamics [see Fig.~\ref{fig:MC}(b)]. Analogous results hold for the case of Abelian weak symmetries.

\subsubsection{General initial states}

For an initial state in a superposition of symmetry eigenstates, $|\psi_0\rangle=\sum_l \mathds{1}_{\mathcal{H}_l}|\psi_0\rangle$, the coherences between symmetry eigenspaces are in general maintained in a quantum trajectory, since in a weakly symmetric representation asymmetric jump operators connect many pairs of symmetry eigenspaces [see Eq.~\eqref{eq:Jj}]. Nevertheless, no new coherences with respect to the eigenvalues of the symmetry superoperator are created, since the action of individual jump operators also features the weak symmetry [cf.~Fig.~\ref{fig:matrix}(b)]. 
Furthermore, for a unique stationary state, from the ergodicity [Eq.~\eqref{eq:ergodic}], any coherences in a quantum trajectory must either interfere destructively in the time average or decay to~$0$ over time. When not only the asymptotic time average in a quantum trajectory but also the asymptotic time-distribution is independent from an initial state, as is the case for generic dynamics~\cite{Benoist2019}, coherences in quantum trajectories necessarily decay to~$0$.

\section{Numerical applications of weakly symmetric representations} \label{sec:numerics}

We now explain how weakly symmetric representations can be used to simplify a construction of the master operator as well as QJMC simulations.

\subsection{Simplified master operator construction}\label{sec:L_cons}

As discussed in Sec.~\ref{sec:L_cons0}, the master operator with a weak symmetry becomes block diagonal in the Liouville representation when using a basis corresponding to symmetry eigenspaces. This simplifies both its diagonalization and numerical integration. We now show how  its construction in such a basis can be further simplified by using a weakly symmetric representation.

\begin{figure}[t!]
	\begin{center}
		\centering
		\includegraphics[width=0.9\columnwidth]{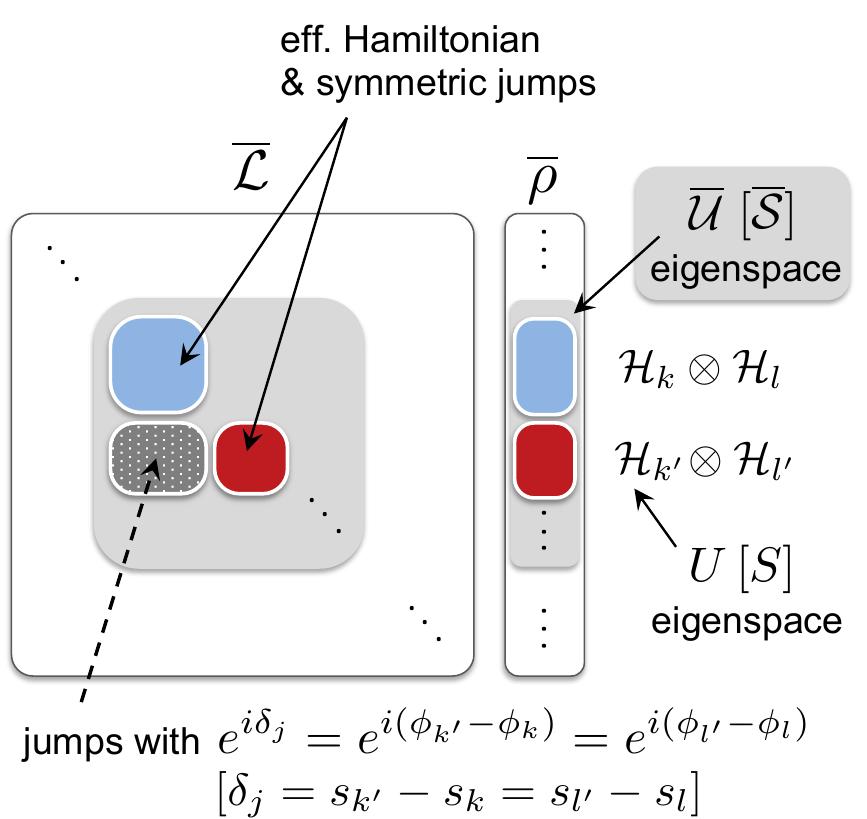} 
		\caption{
			\textbf{Master operator construction with a weakly symmetric representation}. 	When a density matrix is expressed as a vector $\overline{\rho}$, a master operator with the weak symmetry becomes a matrix $\overline{\Lcal}$ [Eq.~\eqref{eq:master_lin}], block diagonal in the eigenspaces of the symmetry $\overline{\Ucal}$ (or $\overline{\Scal}$). 
			A subspace $\mathcal{H}_k\otimes\mathcal{H}_l$ is transformed onto itself by the effective Hamiltonian and symmetric jump operators, while weakly symmetric jump operators connect it to the subspace $\mathcal{H}_{k'}\otimes\mathcal{H}_{l'}$ determined by their $\Ucal$ eigenvalue $e^{i\delta_j}$ (or $\Scal$ eigenvalue $\delta_j$); see~Eqs.~\eqref{eq:master_lin_kl} and~\eqref{eq:master_lin_kl2}. 
			For a discrete weak symmetry with $\mathcal{U}$ with trivially degenerate eigenvalue ratios (or a continuous weak symmetry with $\mathcal{S}$ with trivially degenerate gaps),  $\mathcal{H}_k\otimes\mathcal{H}_l$ with
			$k\neq l$ is always mapped onto itself [dotted gray block vanishes; cf.~Fig.~\ref{fig:matrix}(b)].
		}\vspace*{-5mm}
		\label{fig:matrix2}
	\end{center}
\end{figure}

We have [cf.~Eqs.~\eqref{eq:master_lin} and~\eqref{eq:Heff}]
\begin{eqnarray}\label{eq:master_lin2}
	\overline{\Lcal}&=&-i H_\text{eff}\otimes \mathds{1}+ i \mathds{1}\otimes H_\text{eff}^*+ \sum_jJ_j \otimes J_j^*.
\end{eqnarray}
For dynamics with a weak symmetry in Eq.~\eqref{eq:weakU}, we consider a weakly symmetric representation in Eq.~\eqref{eq:weakR}. From  Eq.~\eqref{eq:HJj}, a subspace $\mathcal{H}_k\otimes\mathcal{H}_l$ is mapped onto itself only by the symmetric effective Hamiltonian and symmetric jump operators (cf.~Fig.~\ref{fig:matrix2}):
\begin{eqnarray}\label{eq:master_lin_kl}
&&(\mathds{1}_{\mathcal{H}_k}\otimes \mathds{1}_{\mathcal{H}_l}^*)\,\overline{\Lcal}\,(\mathds{1}_{\mathcal{H}_k}\otimes \mathds{1}_{\mathcal{H}_l}^*)=\\\nonumber
&& \qquad\qquad-i \mathds{1}_{\mathcal{H}_k} \Tilde{H}_\text{eff} \mathds{1}_{\mathcal{H}_k} \otimes \mathds{1}_{\mathcal{H}_l}^*+ i \mathds{1}_{\mathcal{H}_k}\otimes (\mathds{1}_{\mathcal{H}_l} \Tilde{H}_\text{eff}\mathds{1}_{\mathcal{H}_l})^*
\\\nonumber
&&
\qquad\qquad+ \sum_{j:\, e^{i\delta_j}=1}\mathds{1}_{\mathcal{H}_k} \Tilde{J}_j\mathds{1}_{\mathcal{H}_k} \otimes (\mathds{1}_{\mathcal{H}_l} \Tilde{J}_j \mathds{1}_{\mathcal{H}_l})^*,
\end{eqnarray}
with $ \mathds{1}_{\mathcal{H}_k} \Tilde{H}_\text{eff} \mathds{1}_{\mathcal{H}_k}=\mathds{1}_{\mathcal{H}_k}\Tilde{H} \mathds{1}_{\mathcal{H}_k}- i\sum_{j:\,e^{i\delta_j}=e^{i(\phi_{k'}-\phi_{k})}} (\mathds{1}_{\mathcal{H}_{k'}} \Tilde{J}_j \mathds{1}_{\mathcal{H}_k})^\dagger  (\mathds{1}_{\mathcal{H}_{k'}} \Tilde{J}_j \mathds{1}_{\mathcal{H}_k} )/2$, where $k'$ is such that $e^{i(\phi_{k'}-\phi_{k})}=e^{i\delta_j}$. Furthermore,  it is mapped onto a different subspace $\mathcal{H}_{k'}\otimes\mathcal{H}_{l'}$ corresponding  to the same eigenvalue of $\overline{\Ucal}$  [that is, for $e^{i(\phi_{k'}-\phi_{l'})}= e^{i(\phi_{k}-\phi_{l})}$] only by the  jump operators $\Tilde{J}_j$ with $e^{i\delta_j}=e^{i(\phi_{k'}-\phi_{k})}=e^{i(\phi_{l'}-\phi_{l})}$ (cf.~Fig.~\ref{fig:matrix2}):
\begin{eqnarray}\label{eq:master_lin_kl2}
&&(\mathds{1}_{\mathcal{H}_{k'}}\otimes \mathds{1}_{\mathcal{H}_{l'}}^*)\,\overline{\Lcal}\,(\mathds{1}_{\mathcal{H}_k}\otimes \mathds{1}_{\mathcal{H}_l}^*) =\\\nonumber &&\qquad\qquad\qquad\qquad\!\!\!\!\!\!\!\!\!\!\sum_{\substack{j:\,e^{i\delta_j}=e^{i(\phi_{k'}-\phi_{k})}\\\,\,\,e^{i\delta_j}=e^{i(\phi_{l'}-\phi_{l})}}}\!\!\!\!\!\!\!\!\!\!\mathds{1}_{\mathcal{H}_{k'}} \Tilde{J}_j\mathds{1}_{\mathcal{H}_{k}} \otimes (\mathds{1}_{\mathcal{H}_{l'}} \Tilde{J}_j \mathds{1}_{\mathcal{H}_{l}})^*.\qquad
\end{eqnarray}

The construction of the master operator in the Liouville representation with a weakly symmetric representation is made \emph{efficient}  in two ways. First, only the nontrivial action of $\Lcal$ within the eigenspaces of the symmetry superoperator $\overline{\Ucal}$  is computed, i.e., only nonzero blocks of $\overline{\Lcal}$ are constructed.\footnote{Note that projecting the master operator with a given representation on eigenspaces of the symmetry superoperator, as in Eqs.~\eqref{eq:master_lin_kl} and~\eqref{eq:master_lin_kl2}, effectively leads  to a weakly symmetric representation constructed in Sec.~\ref{sec:R_con_DD}.}
Second, transformations between individual subspaces  $\mathcal{H}_k\otimes\mathcal{H}_l$ can be computed using only the operators in the representation that correspond to a specific eigenvalue of $\Ucal$; see Fig.~\ref{fig:matrix2}.
Therefore, for a minimal symmetric representation, the maximal number of jump operators that contributes in Eq.~\eqref{eq:master_lin_kl} is the maximal number of symmetric jump operators, $\sum_k \dim(\mathcal{H}_k)^2-1$, while in Eq.~\eqref{eq:master_lin_kl2} it is the maximal number of jump operators with the eigenvalue $e^{i(\phi_{k'}-\phi_{k})}$, i.e., $\sum_{l,l':\, e^{i(\phi_{l'}-\phi_{l})}=e^{i(\phi_{k'}-\phi_{k})}} \dim(\mathcal{H}_{l'}\!)\dim(\mathcal{H}_l)$ (as can be seen, for example, by using the Choi representation~\cite{Wolf2012}).

The simplified construction of the master operator is a direct consequence of the simplified structure of quantum trajectories discussed in Sec.~\ref{sec:trajectories}. Indeed, an infinitesimal change in the average system state is a consequence of a change in quantum trajectories either due to the symmetric effective Hamiltonian [cf.~Eq.~\eqref{eq:Upsi_t}] or to an occurrence of a jump $\Tilde{J}_j$ [Eq.~\eqref{eq:UJpsi_t}]. The mapping of $\mathcal{H}_k\otimes\mathcal{H}_l$  for $k=l$ is then determined by quantum trajectories originating in $\mathcal{H}_k$, while for $k\neq l$ it is determined by quantum trajectories of superpositions in $\mathcal{H}_k\oplus\mathcal{H}_l$ (cf.~Fig.~\ref{fig:matrix2}).

Analogous results hold for Abelian weak symmetries. In particular, for a continuous weak symmetry in Eq.~\eqref{eq:weakS} and a weakly symmetric representation in Eq.~\eqref{eq:weakRC}, we have Eq.~\eqref{eq:master_lin_kl} with $e^{i\delta_j}=1$ replaced by $\delta_j=0$, and Eq.~\eqref{eq:master_lin_kl2} with $e^{i\delta_j}=e^{i(\phi_{k'}-\phi_{k})}$ replaced by $\delta_j=s_{k'}-s_k$ for $\mathcal{H}_{k}\otimes\mathcal{H}_{l}$ is mapped onto $\mathcal{H}_{k'}\otimes\mathcal{H}_{l'}$ within the same eigenspace of $\overline{\Scal}$ (that is, for $s_{k'}-s_k= s_{l'}-s_l$).

\subsection{Simplified QJMC simulations}\label{sec:MC}

The QJMC approach (see, e.g., Refs.~\cite{Dum1992,Molmer92,Molmer93,Plenio1998,Daley2014})  is used to generate quantum trajectories in order to obtain dynamics of the average system state via the empirical mean [cf.~Eq.~\eqref{eq:rho_series}] or stationary states of the system [by considering trajectories longer than the relaxation time, or, in the case of a unique stationary state, via Eq.~\eqref{eq:ergodic}]. The advantage of this method in comparison with solving Eq.~\eqref{eq:master}  lies in considering linear operators on pure states in the system space $\mathcal{H}$ rather than the master operator on density matrices in the space isomorphic to $\mathcal{H}\otimes \mathcal{H}$ (cf.~Fig.~\ref{fig:matrix}). 
Since, by definition, the average dynamics governed by the master operator in Eq.~\eqref{eq:master} does not depend on its representation, it can be chosen to simplify QJMC simulations. 
Here, we discuss how  for the dynamics with a weak symmetry, weakly symmetric representations can be utilized (see also Refs.~\cite{Molmer93,Daley2009,Daley2014}).

\subsubsection{Algorithm}

 In order to construct a quantum trajectory up to a finite time $t$, each step consists of two parts. 
First, for an initial state $|\psi_0 \rangle$, time $t_1$  of the first jump is found by drawing a random \emph{uniformly distributed} number  $u_1\in [0,1]$, which represents the probability of no jump occurring until $t_1$ [cf.~Eq.~\eqref{eq:p_t0}],
\begin{equation} \label{eq:u}
u_1=\langle\psi_{t_1} |\psi_{t_1}\rangle=\langle \psi_{0}|e^{i t_1 H_\text{eff}^\dagger} e^{-i t_1 H_\text{eff}}|\psi_{0}\rangle.
\end{equation}

Second, if $t_1> t$, the normalized quantum trajectory at time $t$  is given by normalized Eq.~\eqref{eq:psi_t0}. Otherwise, a jump takes place of the type $j_1$ drawn with the probability $p_{j_1}$ proportional to its instant rate, 
\begin{equation}\label{eq:p_j}
p_j\propto \langle\psi_{t_1}| J_{j}^\dagger J_{j} |\psi_{t_1}\rangle,
\end{equation}
and the system state is updated as $|\psi_{t_1+d t_1}(j_1,t_1)\rangle =  J_{j_1} |\psi_{t_1}\rangle$ [cf.~Eq.~\eqref{eq:psi_t}]. Then, in order to find time $t_2$ of the next jump, the step is repeated with $|\psi_0 \rangle$ replaced by the normalized conditional state $|\psi_{t_1+d t_1}(j_1,t_1)\rangle/\lVert |\psi_{t_1+d t_1}(j_1,t_1)\rangle\rVert$, and $t_1$ replaced by time  $t_2-t_1$ between the first and the second jumps. This is done until $t_{n+1}>t$, in which case  the normalized quantum trajectory after the occurrence of $n$ jumps is given  at time $t$ by normalized Eq.~\eqref{eq:psi_t}. \\

The main computational difficulty  is the evaluation of jump occurrence  times [Eq.~\eqref{eq:u}], which requires finding a norm of a system state evolving under the non-Hermitian effective Hamiltonian $H_\text{eff}$. Therefore, efficient computation of the no-jump dynamics is necessary, e.g., via the exact diagonalization of $H_\text{eff}$  or via fourth-order Runge-Kutta integration~\cite{Plenio1998,Daley2014}. 
We note that, alternatively, a discretization of quantum trajectories to finite but small time steps $\delta t$ can be considered. Here, at each step, occurrence of a jump is decided when a random uniformly distributed number $[0,1]$ is smaller than $ \delta t \sum_{j}\langle\psi| J_{j}^\dagger J_{j} |\psi\rangle$, which approximates,  up to the linear order, the probability of a jump occurring for the system in $|\psi\rangle$ at the beginning of the step [cf.~Eq.~\eqref{eq:u}], upon which the type of jump $j$ is drawn according to its rate [cf.~Eq.~\eqref{eq:p_j}]. The system state is then updated to $J_j|\psi\rangle$  or,  when no jump occurs, to  $(1-i \delta t H_\text{eff})|\psi\rangle$, and subsequently normalized before the next step. This originally proposed approach~\cite{Molmer92,Molmer93} is equivalent to first-order Euler integration of  Eq.~\eqref{eq:u}~\cite{Plenio1998}. 
Nevertheless, another important factor remains: manipulating the conditional system state generally described by $\mathrm{dim}(\mathcal{H})$ coefficients [for example, in Eq.~\eqref{eq:p_j} or when updating and normalizing the state].

\subsubsection{Simplified algorithm}\label{sec:MC0}
\begin{figure}[t]
	\begin{center}
		\includegraphics[width=\columnwidth]{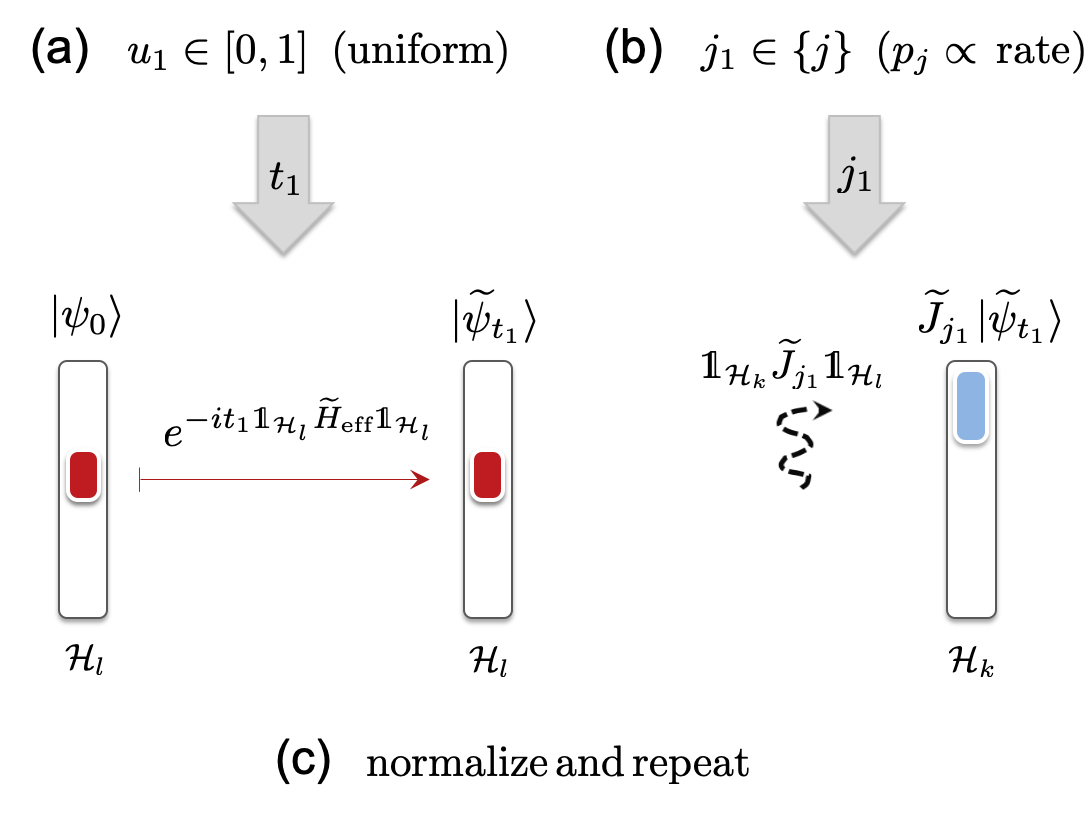} 
		\caption{
			\textbf{Simplified QJMC algorithm}. \textbf{(a)} For an initial symmetric state in an eigenspace $\mathcal{H}_l$ (red), the time $t_1$ of first jump is found by drawing a random $u_1$ that equals the probability of no jumps occurring until $t_1$ determined by the square norm of  $|\Tilde{\psi}_{t_1}\rangle$ evolving with $\mathds{1}_{\mathcal{H}_l} H_\text{eff}\mathds{1}_{\mathcal{H}_l}$  [cf.~Eq.~\eqref{eq:u}]. If $t_1>t$, the system state at time $t$ is chosen as the normalized $|\Tilde{\psi}_{t}\rangle$.
			\textbf{(b)} Otherwise, the type of the first jump is drawn with the probability proportional to the  rate of jump operators $\mathds{1}_{\mathcal{H}_k}\Tilde{J}_{j} \mathds{1}_{\mathcal{H}_l}$, where $k$ corresponds to $\mathcal{H}_k$ with the eigenvalue $e^{i\phi_k}= e^{i\delta_{j_1}}e^{i\phi_l}$, where the updated state $\Tilde{J}_{j}|\Tilde{\psi}_{t_1}\rangle$ is found [cf.~Eq.~\eqref{eq:p_j}]. \textbf{(c)} This state is normalized and used as the initial state in the next step, which is repeated until $t_{n+1}>t$ when the state at time $t$ is chosen as normalized Eq.~\eqref{eq:psi_t}, inside the symmetry eigenspace with the eigenvalue $e^{i\delta_{j_n}}\cdots e^{i\delta_{j_1}} e^{i\phi_l}$.
		}\vspace*{-5mm}
		\label{fig:MC_step}
	\end{center}
\end{figure}

We now explain how the complexity of the QJMC algorithm can be lowered thanks to a weak symmetry by considering a weakly symmetric representation. Simulations are simplified in particular for symmetric initial states, which is relevant for system dynamics with a unique stationary state and for averages of symmetric system observables (see Fig.~\ref{fig:MC_step}). \\

\emph{Symmetric initial states}. For a symmetric initial state, any quantum trajectory is confined to only a single symmetry subspace at a time [cf.~Eq.~\eqref{eq:Upsi_t2} and see Fig.~\ref{fig:MC}(a)]. Therefore, the system state is described at any time  by at most $\max_k\mathrm{dim}(\mathcal{H}_k)$ coefficients (and the label $k$ for the  occupied eigenspace $\mathcal{H}_k$). Furthermore, the evolution with the symmetric effective Hamiltonian in each step of the algorithm can be integrated, if needed, solely on the currently occupied symmetry subspace  [that is, for $\mathcal{H}_l$, $\mathds{1}_{\mathcal{H}_l} \Tilde{H}_\text{eff} \mathds{1}_{\mathcal{H}_l}$  in Eq.~\eqref{eq:u}]. Furthermore, the stochastic dynamics remains exactly the same upon replacing each of the jump operators $J_j$ with the set $\{\mathds{1}_{\mathcal{H}_k} \Tilde{J}_j \mathds{1}_{\mathcal{H}_l}\}_{k,l:\,e^{i\delta_j}=e^{i(\phi_k-\phi_l)}}$ [cf.~Eq.~\eqref{eq:UJpsi_t} and Fig.~\ref{fig:MC}], with the rates of  jumps for a state in $\mathcal{H}_l$ simplified as $\mathds{1}_{\mathcal{H}_l} \Tilde J_{j}^\dagger \Tilde J_{j} \mathds{1}_{\mathcal{H}_l}=(\mathds{1}_{\mathcal{H}_k} \Tilde J_{j} \mathds{1}_{\mathcal{H}_l} )^\dagger (\mathds{1}_{\mathcal{H}_k} \Tilde J_{j} \mathds{1}_{\mathcal{H}_l})$ [for $k$ such that $e^{i\delta_j}=e^{i(\phi_k-\phi_l)}$; cf.~Eq.~\eqref{eq:p_j}]. 

While these improvements are known for the open quantum dynamics microscopically defined by a weakly symmetric representation corresponding to a continuous weak symmetry~\cite{Daley2014}, such as particle losses in a many-body system corresponding to the $U(1)$ symmetry generated by the particle number~\cite{Daley2009}, in this work we show how to implement them for any weak symmetry by constructing a weakly symmetric representation. 
\\

 \emph{General initial states}. For a general initial state, $|\psi_0\rangle\!=\!\sum_l \!\mathds{1}_{\mathcal{H}_l}|\psi_0\rangle$, in each step of the algorithm, the effective Hamiltonian can be integrated, if needed, independently in each symmetry subspace, $|\Tilde \psi_t\rangle\!=\!\sum_l e^{-it\mathds{1}_{\mathcal{H}_l}\! \Tilde{H}_\text{eff}\mathds{1}_{\mathcal{H}_l} }\mathds{1}_{\mathcal{H}_l}|\psi_0\rangle\equiv \sum_{l}\!|\Tilde \psi_t^{(l)}\rangle$ and $\langle\Tilde \psi_t |\Tilde\psi_t\rangle\!=\!\sum_l\langle\Tilde \psi_t^{(l)}|\Tilde \psi_t^{(l)}\rangle$ [cf.~Eq.~\eqref{eq:u}]. Furthermore, a type of occurring jump can be determined with respect to the sum of its rates in individual subspaces, $\langle\Tilde\psi_{t}| \Tilde J _{j}^\dagger \Tilde J_{j} |\Tilde \psi_{t}\rangle=\sum_l  \langle\Tilde \psi_{t}^{(l)}|    (\mathds{1}_{\mathcal{H}_{k(l)}} \Tilde J_{j} \mathds{1}_{\mathcal{H}_l} )^\dagger (\mathds{1}_{\mathcal{H}_k} \Tilde J_{j} \mathds{1}_{\mathcal{H}_l})|\Tilde\psi_{t}^{(l)}\rangle$ [cf.~Eq.~\eqref{eq:p_j}], and the updated state corresponds to the superposition of the updated states, $\sum_l\!  \,\mathds{1}_{\mathcal{H}_{k(l)}}  \Tilde J_{j} \mathds{1}_{\mathcal{H}_l}  |\Tilde\psi_{t}^{(l)}\rangle$, where $k(l)$ is such that $e^{i\delta_j}=e^{i[\phi_{k(l)}-\phi_l]}$.\\

We conclude that QJMC simulations with weakly symmetric representations are simplified in a way comparable to the strong symmetry case with a Hamiltonian and all jump operators being symmetric~\cite{Buca2012,Albert2014}. Indeed, in that case there exists a stationary state $\rho_\text{ss}^{(k)}$ inside each symmetry eigenspace $\mathcal{H}_k$, which can be obtained from  quantum trajectories for an initial state within that subspace  evolving with the effective Hamiltonian $\mathds{1}_{\mathcal{H}_k} H \mathds{1}_{\mathcal{H}_k}$ and jump operators $\{\mathds{1}_{\mathcal{H}_k} J_j \mathds{1}_{\mathcal{H}_k}\}_j$. Similarly, for a general initial state, the dynamics can be solved independently in each symmetry eigenspace.

\subsection{Sparsity}

Hamiltonians and jump operators for many-body system involving only few-body terms are \emph{sparse} in any basis composed as a tensor product of local bases. Since the number of few-body operators scales linearly in the system size, rather than exponentially as the dimension of the system Hilbert space, the master operator is also sparse [cf.~Eq.~\eqref{eq:master_lin}]. This allows for a significant computational speedup in its diagonalization or numerical integration. 

\begin{figure}[t!]
	\begin{center}
		\includegraphics[width=\columnwidth]{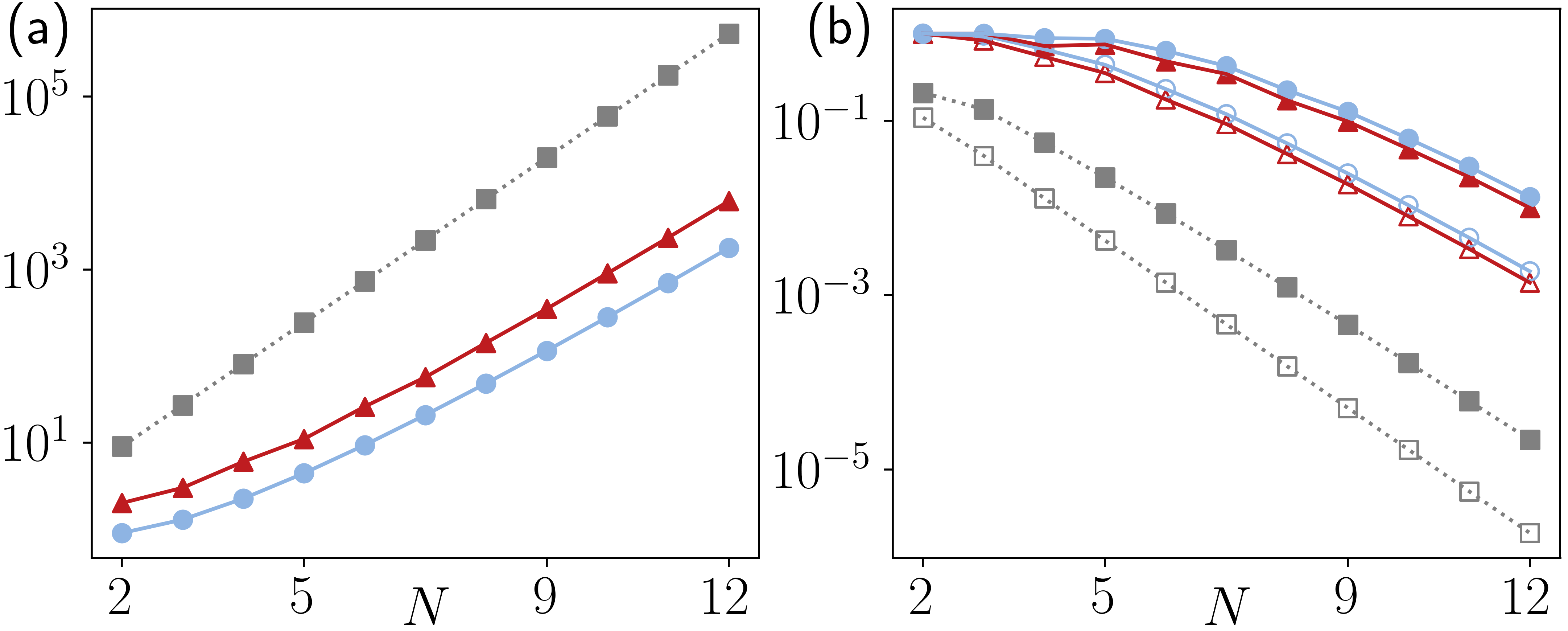} 
		\caption{
				\textbf{Effective dimension and sparsity of a weakly symmetric representation} for dynamics of $N$ spins~$1$ with translation and rotation symmetries.
			\textbf{(a)} Both the average  ({\large\color{Blue}$\bullet$}) and the maximal ({\color{Red}$\blacktriangle$}) dimension among symmetry subspaces, indexed by quasimomentum and total spin along an axis, scales exponentially with $N$ but at a lower rate than the dimension $3^N$ of the system Hilbert space ({\scriptsize \color{Gray}$\blacksquare$}). \textbf{(b)} With nearest-neighbor interactions, $H=\sum_{j=1}^N V \vec{S}_{j}\!\cdot \!\vec {S}_{j+1}$, and local dissipation, $J_{j,\alpha}=\sqrt{\lambda}S_{\alpha}^{(j)}$, $j=1,...,N$, $\alpha=x,y,z$, for the effective Hamiltonian the average sparsity ({\large\color{Blue}$\bullet$}) over the symmetry eigenspaces weighted by their dimension is dominated by the sparsity in the maximal subspace [{\color{Red}$\blacktriangle$}; see also Fig.~\ref{fig:example_trajectories}(d)], and significantly larger than the overall sparsity in the initial basis ({\scriptsize \color{Gray}$\blacksquare$}); cf. Sec.~\ref{sec:examplesT}. Nevertheless, the shown ratios of nonzero to all entries in considered matrices decay exponentially with $N$. Similarly,  the maximum values  among collective jumps,  $\tilde{J}_{q,z}$, $\tilde{J}_{q,+}$, $\tilde{J}_{q,-}$, $q=0,...,N-1$, of the sparsity averaged over blocks connecting pairs of symmetry subspaces [{\large\color{Blue}$\circ$}; with weights given by dimensions of initial subspaces] and the sparsity in the block connecting from the maximal subspace [{\scriptsize \color{Red}$\triangle$}, cf. Figs.~\ref{fig:example_trajectories}(e) and~\ref{fig:example_trajectories}(f)] are larger than the sparsity of local jumps $J_{j,\alpha}$ in the initial basis ({\scriptsize \color{Gray}$\square$}) but decay exponentially with $N$.
		}\vspace*{-7mm}
		\label{fig:example_scaling}
	\end{center}
\end{figure}
\begin{figure*}[t!]
	\begin{center}
		\includegraphics[width=\textwidth]{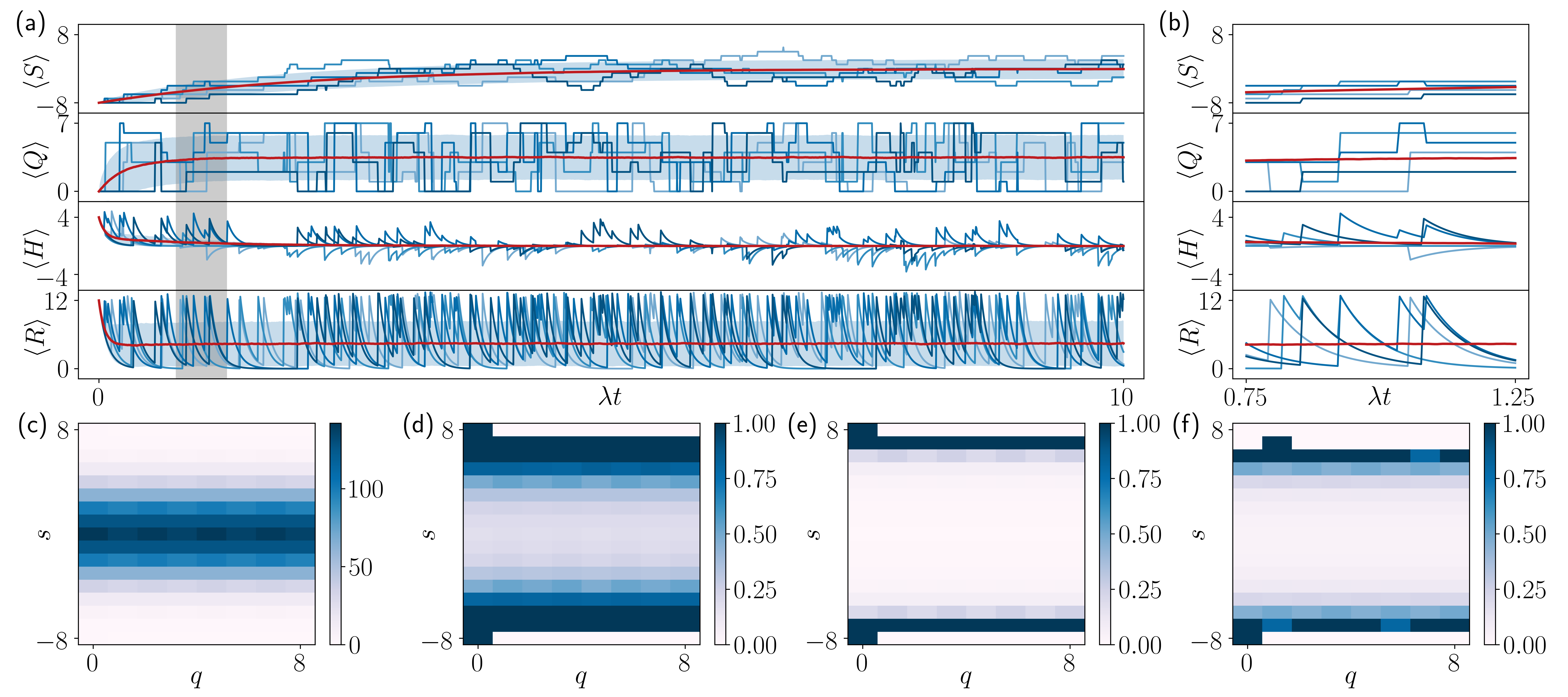} 
		\caption{
				\textbf{QJMC simulations with a weakly symmetric representation} considered in Fig.~\ref{fig:example_scaling} for $N=8$ spins.
			\textbf{(a)} Normalized trajectories for the initial state $S_{z}|\psi_0\rangle=-N|\psi_0\rangle$ and $V/\lambda=1$, shown in terms of average values of (top to bottom) the rotation generator $S=S_z$, the quasimomentum $Q=\sum_{q=0}^{N-1} q \mathds{1}_{\mathcal{H}_q}$, the system Hamiltonian $H$, and the total instant jump rate $R$ [cf. Eq.~\eqref{eq:p_j}]. 
			The mean and fluctuations per trajectory obtained from $10^4$ trajectories are indicated by red curve and blue shading, respectively. 
			\textbf{(b)} Individual trajectories (blue) for the time interval marked by gray in the panel (a) occupy a single symmetry eigenspace at a time (cf. Fig.~\ref{fig:MC}). 
			\textbf{(c)} Dimension of symmetry subspaces $\mathcal{H}_{q,s}$  indexed by quasimomenta $q$ and a total spin $s$ along the  $z$ axis is maximal for $s=0$. 
			\textbf{(d)}-\textbf{(f)} Sparsity in a symmetric basis quantified as the ratio of nonzero entries to all entries in matrices projected onto symmetry subspaces for $H_\text{eff}$~[\textbf{(d)}] and $\tilde{J}_{0,z}$~[\textbf{(e)}], and in matrices projected onto pairs of symmetry subspaces connected by $\tilde{J}_{1,+}$~[\textbf{(f)}; indexed by the initial subspace], is lowest in highly dimensional subspaces [cf. panel (c)]. 
		}\vspace*{-5mm}
		\label{fig:example_trajectories}
	\end{center}
\end{figure*}

Although a weakly symmetric representation features a Hamiltonian and jump operators which are linear combinations of the original operators (cf.~Secs.~\ref{sec:R_con} and~\ref{sec:R_non}), their number scales linearly with system size and thus does not significantly affect the sparsity. In order to exploit the weak symmetry, either for construction and diagonalization of the master operator or for QJMC simulations, however, it is necessary to work in the basis of eigenspaces of the symmetry operator $U$ (cf.~Figs.~\ref{fig:matrix},~\ref{fig:matrix2}, and~\ref{fig:MC}). For a \emph{local symmetry} ($U$ being a tensor product of local unitaries), e.g., a global rotation of spin systems, the basis of symmetry eigenspaces can be chosen separable. For a \emph{nonlocal symmetry}, such as a translation symmetry, symmetric states are entangled.   Therefore, similarly as for symmetries in closed quantum dynamics, it needs to be judged on a case-by-case basis whether exploiting a weak symmetry leads to improved numerical simulations of the open quantum dynamics (see Fig.~\ref{fig:example_scaling} and Sec.~\ref{sec:examples}).

\section{Examples}\label{sec:examples}

Finally, we give closed formulas for weakly symmetric representations of many-body open quantum system dynamics with translation and rotation symmetries. We utilize these representations for numerical simulations of a spin-$1$ chain with nearest-neighbor interactions and local dissipation shown in Figs.~\ref{fig:example_scaling} and~\ref{fig:example_trajectories}, which features both symmetries.

\subsection{Translation symmetry} \label{sec:examplesT}
%

Consider the system composed of $N$ identical subsystems with the Hamiltonian $H= \sum_{j=1}^N  H_j$, where 
$H_j$ is a Hamiltonian for the $j$th subsystem (possibly including interactions). With periodic boundary conditions and in one spacial dimension, the Hamiltonian is translationally invariant, $\mathcal{T} (H_j)=H_{j+1}$, where $\mathcal{T}(\cdot)= T(\cdot)T^\dagger$ and $T|\psi_1\rangle \otimes ...\otimes |\psi_{N-1}\rangle\otimes|\psi_N\rangle=|\psi_N\rangle\otimes|\psi_1\rangle \otimes ...\otimes |\psi_{N-1}\rangle$ so that $T^N=\mathds{1}$. 
For \emph{uniform dissipation}, $J_{j,\alpha}=\mathcal{T}^j (J_{N,\alpha})$, where $j=1,...,N$ and $\alpha$ describes the type of dissipation, the master operators in Eq.~\eqref{eq:master} features the \emph{weak translation symmetry}, $\mathcal{T}\Lcal\mathcal{T}^\dagger=\Lcal$. 

To construct a weakly symmetric representation we can consider the symmetric Hamiltonian $H$ and introduce \emph{collective plane-wave jump} operators,
\begin{equation} \label{eq:Jwave}
\Tilde{J}_{q,\alpha}= \frac{1}{\sqrt{N}} \sum_{j=1}^N e^{-i  j \frac{2\pi q}{N}} J_{j,\alpha}, \quad q=1,...,N,
\end{equation}
where $q$ denotes a quasimomentum and  $\mathcal{T}( \Tilde{J}_{q,\alpha} )= e^{i  \frac{2\pi q}{N}}\Tilde{J}_{q,\alpha}$ [cf.~Eq.~\eqref{eq:weakR}]. 
Equation~\eqref{eq:Jwave} leaves the master operator unchanged as a Fourier transform (a unitary transformation) of the jump operators. For a single type of jump operator (a redundant index $\alpha$), asymmetric jump operators (with $q<N$) in Eq.~\eqref{eq:Jwave}  are uniquely defined  (cf.~Sec.~\ref{sec:R_non}). 

A symmetric basis can be determined as follows. For a local basis of the system chosen as product states of identical subsystem bases, each element belongs to a cycle of a length $l$ under the translation symmetry, where $l$ divides $N$. The plane-wave superpositions of the basis elements with quasimomenta $q_l$ corresponding to that cycle are  eigenstates of $T$ corresponding to eigenvalues  $e^{i 2\pi q_l /l}$, where $q_l=0,1,...,l-1$, so that the effective quasimomenta is $q=N q_l /l$. Therefore the dimension of a symmetry subspace indexed by $q$ equals the number of cycles with lengths $l= N/d_q$, where $d_q$ is a common divisor of $q$ and $N$ [cf.~Figs.~\ref{fig:example_scaling}(a) and~\ref{fig:example_trajectories}(c)].

For \emph{local interactions and dissipation}, the effective Hamiltonian and collective jump operators in Eq.~\eqref{eq:Jwave} remain \emph{sparse} in the translationally symmetric basis constructed above [cf.~Figs.~\ref{fig:example_scaling}(b),~\ref{fig:example_scaling}(c), and~\ref{fig:example_trajectories}(d)-\ref{fig:example_trajectories}(f)]. Indeed, the basis elements are superpositions of at most $N$ local states. 
	Thus, in such a basis, a matrix for the effective Hamiltonian  features at most $N^2 z$ nonzero elements in each column, where $z$ is the maximal number of nonzero entries in columns of the matrix in the local basis. 
	Similarly, collective plane-wave jump operators in Eq.~\eqref{eq:Jwave} being linear combinations of $N$ jump operators feature at most $N^3 z_\alpha$ nonzero elements in each column, where $z_\alpha$ is the maximal number of nonzero entries in columns of the matrix in the local basis for a jump of type~$\alpha$. For a local Hamiltonian $H$ and local jumps $J_{j,\alpha}$, $z$ and $z_\alpha$ are independent from $N$.


\subsection{Rotation symmetry} 

Consider the system of $N$ spins with the Hamiltonian $H= \sum_{jk=1}^N V_{jk}\, \vec{S}_j\!\cdot\! \vec{S}_k+\sum_{jklm=1}^N W_{jklm}\, (\vec{S}_j\!\cdot\! \vec{S}_k)(\vec{S}_l\!\cdot\! \vec{S}_m)$, where $\vec S_j=[S_x^{(j)},S_y^{(j)},S_z^{(j)}]$, with $S_\alpha^{(j)}$ being the $j$th spin operator for $\alpha$ direction, while the dissipation corresponds to \emph{local depolarization}, $J_{j,\alpha}=\sqrt{\lambda_{j}}\, S_\alpha^{(k)}$, with $j=1,...,N$ and $\alpha=x,y,z$. The dynamics features the \emph{weak rotation symmetry}, as  $[\vec{n}\cdot\vec{\Scal},\Lcal]=0$, where $\vec {\Scal}=(\Scal_x,\Scal_y,\Scal_z)$ is a vector of generators of rotation of all spins around the axes, with $S_\alpha=\sum_{j=1}^N  S_\alpha^{(j)}$ for  $\alpha=x,y,z$   and $\vec{n}\in \mathbb{R}^3$ [cf.~Eq.~\eqref{eq:weakS}]. 

A weakly symmetric representation for $U(1)$ symmetry generated by $S_z$ can be obtained by keeping the Hamiltonian $H$, while replacing $J_{j,x}$ and $J_{j,y}$ with $J_{j,\pm}= (J_{j,x}\pm i J_{j,y})/\sqrt{2}$, which, respectively, increase and lower $S_z^{(j)}$ by $1$,  and thus $\Scal_z(J_{j,\pm})
=\pm {J}_{j,\pm}$ [cf.~Eq.~\eqref{eq:weakRC} and see Figs.~\ref{fig:example_trajectories}(a) and~\ref{fig:example_trajectories}(b)]. These operators act on the symmetry subspaces composed of parts corresponding to fixed numbers of spins with a given spin value along the $z$-axis, whose dimension is given by multinomial coefficients [cf.~Fig.~\ref{fig:example_trajectories}(c)].
Analogous constructions hold for $S_x$ and $S_y$, but the representations differ, as these generators do not commute with $S_z$. In contrast, when the weak translation symmetry is also present, e.g., $\lambda_j\equiv\lambda$, $V_{jk}\equiv V_{|j-k|}$, and $W_{jklm}\equiv W_{|j-k| |l-m|}$ (see Fig.~\ref{fig:example_trajectories}), the jump operators in Eq.~\eqref{eq:Jwave} for $\alpha=z,+,-$, are simultaneously eigenmatrices of $\Scal_z$ and $\mathcal{T}$ (note that $[S_z,T]=0$). Note that  the  jump operators $\tilde{J}_{q,\pm}$ with $q<N$  are uniquely determined (cf.~Sec.~\ref{sec:R_non}).

\section{Conclusions}\label{sec:conclusions}

In this article, we investigated how Abelian weak symmetries in Markovian dynamics of open quantum systems can be translated into their quantum stochastic dynamics. We showed how to construct weakly symmetric representations of a master operator governing the dynamics, for which quantum trajectories are symmetric, i.e., are found within a single symmetry eigenspace at a time, whenever initial system states are chosen symmetric. This enabled us to exploit weak symmetries of the dynamics for simplifying the QJMC algorithm, with the memory and processing required for simulations reduced in a way akin to the case of strong symmetries. We also demonstrated how the efficiency in constructing the Liouville representation of the master operator can be improved -- a result directly relevant for solving the system evolution via diagonalization or numerical integration of the master operator.  Finally, we note that the existence of weakly symmetric representations does not rely on the dynamics being time independent, as considered, for simplicity, in this work. For a time-dependent dynamics that features the same weak symmetry at all times, a time-dependent weakly symmetric representation can be analogously constructed from given time-dependent Hamiltonian and jump operators.

\section*{Acknowledgements}

K.~M. thanks E.~I.~Rodriguez Chiacchio for useful discussions and gratefully acknowledges the support from a Henslow Research Fellowship.   D.~C.~R. acknowledges support from the University of Nottingham through grant no.\ FiF1/3.

\bibliography{symmetry.bib}

\end{document}